\documentclass[english,10pt,twocolumn,twoside]{IEEEtran}
\usepackage[T1]{fontenc}
\usepackage[latin9]{inputenc}
\usepackage{bm}
\usepackage{amsthm}
\usepackage{amsmath}
\usepackage{amssymb}
\usepackage{graphicx}

\makeatletter
\theoremstyle{plain}
\newtheorem{thm}{\protect\theoremname}
\theoremstyle{plain}
\newtheorem{prop}[thm]{\protect\propositionname}

\usepackage{colortbl}

\makeatother

\usepackage{babel}
\providecommand{\propositionname}{Proposition}
\providecommand{\theoremname}{Theorem}

\begin{document}

\title{A Unifying Framework for Adaptive\linebreak{}
 Radar Detection in Homogeneous\linebreak{}
 plus Structured Interference-Part I:\linebreak{}
 On the Maximal Invariant Statistic}

\author{D.~Ciuonzo,\IEEEmembership{~Member,~IEEE,} A.~De~Maio,~\IEEEmembership{Fellow,~IEEE,}
and D. Orlando, \IEEEmembership{Senior~Member,~IEEE}\thanks{Manuscript received 17th July 2015.\protect \\
D. Ciuonzo and A. De Maio are with University of Naples \textquotedbl{}Federico
II\textquotedbl{}, DIETI, Via Claudio 21, 80125 Naples, Italy. E-mail:
domenico.ciuonzo@ieee.org; ademaio@unina.it. \protect \\
D. Orlando is with Università degli Studi \textquotedblleft Niccolò
Cusano\textquotedblright , Via Don Carlo Gnocchi 3, 00166 Roma, Italy.
E-mail: danilo.orlando@unicusano.it.\protect \\
}}
\maketitle
\begin{abstract}
This paper deals with the problem of adaptive multidimensional/multichannel
signal detection in homogeneous Gaussian disturbance with unknown
covariance matrix and structured deterministic interference. The aforementioned
problem corresponds to a generalization of the well-known Generalized
Multivariate Analysis of Variance (GMANOVA). In this first part of
the work, we formulate the considered problem in canonical form and,
after identifying a desirable group of transformations for the considered
hypothesis testing, we derive a Maximal Invariant Statistic (MIS)
for the problem at hand. Furthermore, we provide the MIS distribution
in the form of a stochastic representation. Finally, strong connections
to the MIS obtained in the open literature in simpler scenarios are
underlined. \end{abstract}


\section{Introduction}

\subsection{Motivation and Related Works}

\IEEEPARstart{T}{he problem} of adaptive detection of targets embedded
in Gaussian interference is an active research field which has been
object of great interest in the last decades. Many works appeared
in the open literature, dealing with the design and performance analysis
of several detectors handling many specific detection problems (the
interested reader is referred to \cite{Gini2001} and references therein). 

It can be shown that most of the aforementioned models can be seen
as special cases of the model considered by Kelly and Forsythe \cite{Kelly1989},
which is very general and encompasses point-like and extended targets
as special instances. The considered model allows for training samples
which contain random interference modeled as an unknown covariance
matrix that accounts for both clutter and thermal noise, with the
implict assumption that signal plus noise and noise-only vector samples
share the same covariance matrix, thus determining a so-called \emph{homogeneous
environment}.

The signal model considered in the aforementioned report is the well-known
Generalized Multivariate Analysis of Variance (GMANOVA) in statistics
literature \cite{Muirhead2009}, also referred to as a ``double-subspace''
signal model (see for example \cite{Liu2014,Liu2014a}). The standard
GMANOVA model was  first formulated by Potthoff and Roy \cite{Potthoff1964}
and  consists in a generic patterned mean problem with a data matrix
whose columns are normal random vectors with a common unknown covariance
matrix. The GMANOVA model was later studied in more detail in \cite{Khatri1966},
where maximum likelihood estimates of unknown parameters were obtained.
For a detailed introduction to estimation and detection in GMANOVA
model (along with few interesting application examples) the interested
reader may refer to the excellent tutorial \cite{Dogandzic2003}.

Differently, in this paper we will study a modified version of GMANOVA
with respect to its classical formulation \cite{Dogandzic2003}, referred
to as I-GMANOVA in what follows. The considered model allows for the
presence of a structured (partially known) non-zero mean under both
hypotheses. Such  disturbance is collectively represented as an unknown
deterministic matrix, which determines  an additional set of nuisance
parameters for the considered hypothesis testing (i.e., other than
the covariance matrix). The aforementioned model easily accounts for
the presence of structured subspace interference affecting the target
detection task. Thus it is clear that taking such interference into
consideration enables the application of this model to adaptive radar
detection; for instance it may accomodate the presence of multiple
pulsed coherent jammers impinging on the radar antenna from some directions.

Although several different decision criteria can be considered to
attack composite hypothesis testing problems \cite{Kay1998,Lehmann2006},
an elegant and systematic way consists in resorting to the so-called
\emph{Principle of Invariance} \cite{Scharf1991,Lehmann2006}. Indeed,
the aforementioned principle\emph{,} when exploited at the design
stage, allows to focus on decision rules enjoying some desirable practical
features. The preliminary step consists in individuating a suitable
group of trasformations which leaves the formal structure of the hypothesis
testing problem unaltered. With reference to radar adaptive detection,
the mentioned principle represents an effective tool for obtaining
a statistic which is invariant with respect to the set of nuisance
parameters, therefore constituting the basis for Constant-False Alarm
Rate (CFAR) rules. Indeed, every invariant decision rule can be written
in terms of the maximal invariant statistic. Therefore, with reference
to I-GMANOVA model, the principle of invariance allows for imposing
CFARness property with respect to the clutter plus noise (disturbance)
covariance matrix and the jammer location parameters. 

It is worth remarking that the use of the invariance principle for
generic composite hypothesis testing problems \cite{Scharf1991,Lehmann2006}
(and, more specifically, in the context of radar adaptive detection)
\emph{is not new}. Indeed, starting from the seminal paper \cite{Bose1995},
many works focused on adaptive radar detection problems with the use
of \emph{invariance theory}. For example, in \cite{Scharf1994,Raghavan1996,Bose1996},
invariance theory was exploited to study the problem of single-subspace
(adaptive) detection of point-like targets. Later, similar works appeared
in the open literature dealing with the case of a target spread among
more range cells \cite{Conte2003,Raghavan2013}. More recently, the
same statistical tool has been employed to address the problem of
adaptive single-subspace detection problem (of point-like targets)
in the joint presence of random and subspace structured interference
in \cite{A.DeMaio2014}. In this respect, we build upon the aforementioned
results in order to develop an exhaustive study for the considered
I-GMANOVA model under the point of view of the invariance.

\subsection{Summary of the contributions and Paper Organization}

The main contributions of the first part of the present study are
summarized as follows:
\begin{itemize}
\item We first show that the problem at hand admits a more intuitive representation,
by exploiting a \emph{canonical form} representation. Such representation
helps obtaining the maximal invariant statistics and  gaining insights
for the problem under investigation;
\item The group of transformations which leaves the problem invariant is
identified, thus allowing the search for a MIS. 
\item Given the aforementioned group of transformations, the canonical form
is exploited in order to obtain the MIS, which, for the I-GMANOVA
model is represented by \emph{two matrices} which compress the original
data. Such result can be interpreted as the generalization of the
two-components scalar MIS obtained in the classical references \cite{Bose1995,Raghavan1996}.
\item A theoretical performance analysis of the MIS is obtained, in terms
of its distribution. Even though in the considered setup the MIS does
not generally admit an explicit  expression for its probability density
function (pdf), a simpler form of the statistic distribution, by means
of a suitable stochastic representation, is provided.
\item Finally, the obtained MIS expression is compared with similar findings
obtained in the literature for simpler  scenarios, thus showing that
the aforementioned cases can be seen as special instances of the obtained
MIS.
\end{itemize}
The explicit expression of the MIS obtained in this first part is
then exploited to show CFARness of all the  detectors considered in
part II of this work.

The remainder of the paper is organized as follows: in Sec. \ref{sec: Problem formulation}
we introduce the hypothesis testing problem under investigation; in
Sec. \ref{sec: MIS} we describe the desirable invariance properties
and derive the MIS. Sec. \ref{sec: Stat Char MIS} is devoted to the
statistical characterization of the MIS, while in Sec. \ref{sec: MIS (special instances)}
we particularize the MIS to specific instances and compare it with
previously obtained results in the open literature. Some concluding
remarks and future research directions are given in Sec. \ref{sec: Conclusions};
finally, proofs and derivations are confined to the Appendices.

\emph{Notation} - Lower-case (resp. Upper-case) bold letters denote
vectors (resp. matrices), with $a_{n}$ (resp. $A_{n,m}$) representing
the $n$th (resp. the $(n,m)$th) element of the vector $\bm{a}$
(resp. matrix $\bm{A}$); $\mathbb{R}^{N}$, $\mathbb{C}^{N}$, and
$\mathbb{H}^{N\times N}$ are the sets of $N$-dimensional vectors
of real numbers, of complex numbers, and of $N\times N$ Hermitian
matrices, respectively; upper-case calligraphic letters and braces
denote finite sets; $\mathbb{E}\{\cdot\}$, $\mathrm{Cov}[\cdot]$,
$(\cdot)^{T}$, $(\cdot)^{\dagger}$, $\angle\cdot$, $\mathrm{Tr}\left[\cdot\right]$,
denote expectation, covariance, transpose, Hermitian, phase and matrix
trace operators, respectively; $\bm{0}_{N\times M}$ (resp. $\bm{I}_{N}$)
denotes the $N\times M$ null (resp. identity) matrix; $\bm{0}_{N}$
(resp. $\bm{1}_{N}$) denotes the null (resp. ones) column vector
of length $N$; $\mathrm{vec}(\bm{M})$ stacks the first to the last
column of the matrix $\bm{M}$ one under another to form a long vector;
$\det(\bm{A})$ and $||\bm{A}||_{F}$ denote the determinant and Frobenius
norm of matrix $\bm{A}$; $\bm{A}\otimes\bm{B}$ indicates the Kronecker
product between $\bm{A}$ and $\bm{B}$ matrices; $\mathrm{diag(}\bm{A},\bm{B})$
denotes the block-diagonal matrix obtained by placing matrices $\bm{A}$
and $\bm{B}$ along the main diagonal; the symbol ``$\sim$'' means
``distributed as''; $\bm{x}\sim\mathcal{C}\mathcal{N}_{N}(\bm{\mu},\bm{\Sigma})$
denotes a complex (proper) Gaussian-distributed vector $\bm{x}$ with
mean vector $\bm{\mu}\in\mathbb{C}^{N\times1}$ and covariance matrix
$\bm{\Sigma}\in\mathbb{C}^{N\times N}$; $\bm{X}\sim\mathcal{C}\mathcal{N}_{N\times M}(\bm{A},\bm{B},\bm{C})$
denotes a complex (proper) Gaussian-distributed matrix $\bm{X}$ with
mean $\bm{A}\in\mathbb{C}^{N\times M}$ and $\mathrm{Cov}[\mathrm{vec}(\bm{X})]=\bm{B}\otimes\bm{C}$;
$\bm{S}\sim\mathcal{C}\mathcal{W}_{N}(K,\,\bm{A})$ denotes a complex
central Wishart distributed matrix $\bm{S}$ with parameters $K\in\mathbb{N}$
and $\bm{A}\in\mathbb{C}^{N\times N}$ positive definite matrix; $\bm{M}\sim\mathcal{CF}_{a}(\bm{A},\ell,m)$
is a non-central multivariate complex F distributed matrix $\bm{M}$
with mean $\bm{A}$ and parameters $a$, $\ell$, and $m$; $\bm{P}_{A}$
denotes the orthogonal projection of the full-column-rank matrix $\bm{A}$,
that is $\bm{P}_{\bm{A}}\triangleq[\bm{A}(\bm{A}^{\dagger}\bm{A})^{-1}\bm{A}^{\dagger}]$,
while $\bm{P}_{A}^{\perp}$ its complement, that is $\bm{P}_{A}^{\perp}\triangleq(\bm{I}-\bm{P}_{\bm{A}})$.

\section{Problem Formulation \label{sec: Problem formulation}}

We assume that a matrix of complex-valued samples $\bm{X}\in\mathbb{C}^{N\times K}$
is collected, accounting for both primary (signal-bearing) and secondary
(signal-free) data. The hypothesis testing problem under investigation
can be formulated as:
\begin{equation}
\begin{cases}
\mathcal{H}_{0}: & \bm{X}=\widetilde{\bm{A}}_{t}\,\widetilde{\bm{B}}_{t}\,\widetilde{\bm{C}}+\bm{N}_{0}\\
\mathcal{H}_{1}: & \bm{X}=\left(\widetilde{\bm{A}}_{t}\,\widetilde{\bm{B}}_{t}+\widetilde{\bm{A}}_{r}\,\widetilde{\bm{B}}_{r}\right)\widetilde{\bm{C}}+\bm{N}_{0}
\end{cases}\label{eq: Hypothesis testing formulation}
\end{equation}

where:
\begin{itemize}
\item $\bm{N}_{0}\in\mathbb{C}^{N\times K}$ is a matrix whose columns are
independent and identically distributed (iid) proper complex normal
random vectors with zero mean and (unknown) positive definite covariance
matrix $\bm{R}_{\star}\in\mathbb{C}^{N\times N}$, that is $\bm{N}_{0}\sim\mathcal{CN}_{N\times K}(\bm{0}_{N\times K},\bm{I}_{K},\bm{R}_{\star})$;
\item $\widetilde{\bm{B}}_{t}\in\mathbb{C}^{t\times M}$ and $\widetilde{\bm{B}}_{r}\in\mathbb{C}^{r\times M}$
denote the (unknown) deterministic matrix coordinates, representing
the interference and the useful signal, respectively;
\item $\widetilde{\bm{A}}_{t}\in\mathbb{C}^{N\times t}$ and $\widetilde{\bm{A}}_{r}\in\mathbb{C}^{N\times r}$
represent the (known) left subspace of the interference and the useful
signal, respectively. The matrices $\widetilde{\bm{A}}_{t}$ and $\widetilde{\bm{A}}_{r}$
are both assumed full-column rank, with their columns being linearly
independent;
\item Similarly, $\widetilde{\bm{C}}\in\mathbb{C}^{M\times K}$ is a known
matrix describing the right subspace associated to \emph{both} signal
and interference; the matrix $\widetilde{\bm{C}}$ is assumed full-row-rank.
\end{itemize}
Additionally, aiming at a compact notation, we define $\widetilde{\bm{A}}\triangleq\begin{bmatrix}\widetilde{\bm{A}}_{t} & \widetilde{\bm{A}}_{r}\end{bmatrix}\in\mathbb{C}^{N\times J}$
and $\widetilde{\bm{B}}\triangleq\begin{bmatrix}\widetilde{\bm{B}}_{t}^{T} & \widetilde{\bm{B}}_{r}^{T}\end{bmatrix}^{T}\in\mathbb{C}^{J\times M}$,
where we have denoted $J\triangleq r+t$.

From inspection of Eq. (\ref{eq: Hypothesis testing formulation}),
we notice that the considered test has a complicated structure, which
is thus difficult to analyze. Therefore, before proceeding further,
we first show that a simpler equivalent formulation of the considered
problem can be obtained in the so-called \emph{canonical form} \cite{Kelly1989}.

With this intent, we first consider the QR-decomposition of $\widetilde{\bm{A}}=\bm{Q}_{\alpha}\bm{R}_{\alpha}$,
where $\bm{Q}_{\alpha}\in\mathbb{C}^{N\times J}$ is a slice of a
unitary matrix (i.e., $\bm{Q}_{\alpha}^{\dagger}\bm{Q}_{\alpha}=\bm{I}_{J}$)
and $\bm{R}_{\alpha}\in\mathbb{C}^{J\times J}$ a non-singular upper
triangular matrix. It can be readily shown that $\bm{Q}_{\alpha}$
and $\bm{R}_{\alpha}$ can be conveniently partitioned as:
\begin{equation}
\bm{Q}_{\alpha}=\begin{bmatrix}\bm{Q}_{\alpha,t} & \bm{Q}_{\alpha,r}\end{bmatrix}\qquad\bm{R}_{\alpha}=\begin{bmatrix}\bm{R}_{\alpha,t} & \bm{R}_{\alpha,x}\\
\bm{0}_{r\times t} & \bm{R}_{\alpha,r}
\end{bmatrix}\label{eq: QR decomp (canonical form)}
\end{equation}
where $\bm{Q}_{\alpha,t}\in\mathbb{C}^{N\times t}$ and $\bm{R}_{\alpha,t}\in\mathbb{C}^{t\times t}$
arise from the QR-decomposition of $\widetilde{\bm{A}}_{t}$, namely
$\widetilde{\bm{A}}_{t}=\bm{Q}_{\alpha,t}\,\bm{R}_{\alpha,t}$, with
$\bm{Q}_{\alpha,t}$ such that $\bm{Q}_{\alpha,t}^{\dagger}\,\bm{Q}_{\alpha,t}=\bm{I}_{t}$
and $\bm{R}_{\alpha,t}$ a non-singular upper triangular matrix. Furthermore,
$\bm{R}_{\alpha,x}\in\mathbb{C}^{t\times r}$, and $\bm{R}_{\alpha,r}\in\mathbb{C}^{r\times r}$
is another non-singular upper triangular matrix. Similarly, $\bm{Q}_{\alpha.r}\in\mathbb{C}^{N\times r}$
is such that $\bm{Q}_{\alpha.r}^{\dagger}\bm{Q}_{\alpha.r}=\bm{I}_{r}$.
Equalities in Eq. (\ref{eq: QR decomp (canonical form)}) are almost
evident consequences of the well-known Gram-Schmidt procedure \cite{Horn2012}.
Now, let us define a unitary matrix $\bm{U}_{\alpha}\in\mathbb{C}^{N\times N}$
whose first $J$ columns are collectively equal to $\bm{Q}_{\alpha}$.
Then, it follows that:
\begin{gather}
\bm{A}\triangleq\underbrace{\bm{U}_{\alpha}^{\dagger}\,\bm{Q}_{\alpha}}_{\in\mathbb{C}^{N\times J}}=\begin{bmatrix}\bm{I}_{t} & \bm{0}_{t\times r}\\
\bm{0}_{r\times t} & \bm{I}_{r}\\
\bm{0}_{(N-J)\times t} & \bm{0}_{(N-J)\times r}
\end{bmatrix}=\begin{bmatrix}\bm{E}_{t} & \bm{E}_{r}\end{bmatrix}\label{eq: rotated QR}
\end{gather}
where $\bm{E}_{t}\triangleq\begin{bmatrix}\bm{I}_{t} & \bm{0}_{t\times r} & \bm{0}_{t\times(N-J)}\end{bmatrix}^{T}$
and $\bm{E}_{r}\triangleq\begin{bmatrix}\bm{0}_{r\times t} & \bm{I}_{r} & \bm{0}_{r\times(N-J)}\end{bmatrix}^{T}$,
respectively. Also, let $\widetilde{\bm{C}}$ be expressed in terms
of its Singular Value Decomposition (SVD) as
\begin{equation}
\widetilde{\bm{C}}=\bm{U}_{\gamma}\,\bm{\Lambda}_{\gamma}\,\bm{V}_{\gamma}^{\dagger}\,,
\end{equation}
where $\bm{U}_{\gamma}\in\mathbb{C}^{M\times M}$ and $\bm{V}_{\gamma}\in\mathbb{C}^{K\times K}$
are both unitary matrices, and the matrix of the singular values $\bm{\Lambda}_{\gamma}\in\mathbb{C}^{M\times K}$
has the following noteworthy form:
\begin{equation}
\bm{\Lambda}_{\gamma}=\begin{bmatrix}\widetilde{\bm{\Lambda}}_{\gamma} & \bm{0}_{M\times(K-M)}\end{bmatrix}\,,
\end{equation}
with $\widetilde{\bm{\Lambda}}_{\gamma}\in\mathbb{C}^{M\times M}$
being a diagonal matrix. Therefore 
\begin{equation}
\widetilde{\bm{C}}\bm{V}_{\gamma}=\bm{M}_{\gamma}\,\begin{bmatrix}\bm{I}_{M} & \bm{0}_{M\times(K-M)}\end{bmatrix}
\end{equation}
 holds, where $\bm{M}_{\gamma}\triangleq\bm{U}_{\gamma}\widetilde{\bm{\Lambda}}_{\gamma}$. 

Given the aforementioned definitions, without loss of generality we
will consider the transformed data matrix $\bm{Z}\triangleq(\bm{U}_{\alpha}^{\dagger}\,\bm{X}\,\bm{V}_{\gamma})\in\mathbb{C}^{N\times K}$
in what follows. Such transformation does not alter the hypothesis
testing problem being considered, as it simply applies left and right
rotations to data matrix $\bm{X}$ (viz. multiplications by unitary
matrices). The new data matrix, when $\mathcal{H}_{1}$ is in force,
can be expressed as:
\begin{align}
\bm{Z} & =\bm{U}_{\alpha}^{\dagger}\left(\bm{Q}_{\alpha}\,\bm{R}_{\alpha}\,\widetilde{\bm{B}}\,\bm{U}_{\gamma}\,\bm{\Lambda}_{\gamma}\,\bm{V}_{\gamma}^{\dagger}\right)\bm{V}_{\gamma}+\bm{N}\\
 & =\bm{A}\,(\bm{R}_{\alpha}\widetilde{\bm{B}}\,\bm{M}_{\gamma})\begin{bmatrix}\bm{I}_{M} & \bm{0}_{M\times(K-M)}\end{bmatrix}+\bm{N}\\
 & =\bm{A}\,\begin{bmatrix}\bm{B}_{t,1}\\
\bm{B}
\end{bmatrix}\,\begin{bmatrix}\bm{I}_{M} & \bm{0}_{M\times(K-M)}\end{bmatrix}+\bm{N}
\end{align}
where we have defined $\bm{B}_{t,1}\triangleq((\bm{R}_{\alpha,t}\,\widetilde{\bm{B}}_{t}+\bm{R}_{\alpha,x}\,\widetilde{\bm{B}}_{r})\,\bm{M}_{\gamma})\in\mathbb{C}^{t\times M}$,
$\bm{B}\triangleq(\bm{R}_{\alpha,r}\,\widetilde{\bm{B}}_{r}\,\bm{M}_{\gamma})\in\mathbb{C}^{r\times M}$
and $\bm{N}\triangleq(\bm{U}_{\alpha}^{\dagger}\,\bm{N}_{0}\,\bm{V}_{\gamma})\in\mathbb{C}^{N\times K}$,
respectively. Furthermore, for the sake of notational convenience,
we define $\bm{B}_{s}\triangleq\begin{bmatrix}\bm{B}_{t,1}^{T} & \bm{B}^{T}\end{bmatrix}^{T}$.
On the other hand, when $\mathcal{H}_{0}$ holds true, the matrix
$\bm{Z}$ can be expressed as:
\begin{align}
\bm{Z} & =\bm{U}_{\alpha}^{\dagger}\left(\bm{Q}_{\alpha}\,\bm{R}_{\alpha}\,\begin{bmatrix}\widetilde{\bm{B}}_{t}\\
\bm{0}_{r\times M}
\end{bmatrix}\,\bm{U}_{\gamma}\,\bm{\Lambda}_{\gamma}\,\bm{V}_{\gamma}^{\dagger}\right)\bm{V}_{\gamma}+\bm{N}\\
 & =\bm{A}\,\begin{bmatrix}\bm{B}_{t,0}\\
\bm{0}_{r\times M}
\end{bmatrix}\,\begin{bmatrix}\bm{I}_{M} & \bm{0}_{M\times(K-M)}\end{bmatrix}+\bm{N}
\end{align}
where $\bm{B}_{t,0}\triangleq(\bm{R}_{\alpha,t}\,\widetilde{\bm{B}}_{t}\,\bm{M}_{\gamma})\in\mathbb{C}^{t\times M}$.
Furthermore, aiming at keeping a compact notation, we will employ
the definition $\bm{C}\triangleq\begin{bmatrix}\bm{I}_{M} & \bm{0}_{M\times(K-M)}\end{bmatrix}$
in what follows. Gathering all the above results, the problem in Eq.
(\ref{eq: Hypothesis testing formulation}) can be equivalently rewritten
in terms of $\bm{Z}$ as:
\begin{equation}
\begin{cases}
\mathcal{H}_{0}: & \bm{Z}=\bm{A}\,\begin{bmatrix}\bm{B}_{t,0}\\
\bm{0}_{r\times M}
\end{bmatrix}\,\bm{C}+\bm{N}\\
\mathcal{H}_{1}: & \bm{Z}=\bm{A}\,\bm{B}_{s}\,\bm{C}+\bm{N}
\end{cases}\label{eq: Transformed data - hypothesis testing problem}
\end{equation}
Finally we recall that, since $\bm{N}_{0}\sim\mathcal{CN}_{N\times K}(\bm{0}_{N\times K},\bm{I}_{K},\bm{R}_{\star})$,
$\bm{N}$ is distributed as $\bm{N}\sim\mathcal{CN}_{N\times K}(\bm{0}_{N\times K},\bm{I}_{K},\bm{R})$,
where $\bm{R}\triangleq(\bm{U}_{\alpha}^{\dagger}\,\bm{R}_{\star}\,\bm{U}_{\alpha})$
\cite{Kelly1989}. 

An important remark is now in order. Specifically, for the problem
in Eq. (\ref{eq: Hypothesis testing formulation}), the relevant parameter
to decide for the presence of a target is $\widetilde{\bm{B}}_{r}$.
Otherwise stated, if the hypothesis $\mathcal{H}_{1}$ holds true,
then $||\widetilde{\bm{B}}_{r}||_{F}>0$, while $||\widetilde{\bm{B}}_{r}||_{F}=0$
under the target-absent hypothesis ($\mathcal{H}_{0}$). As a consequence,
since $\bm{R}_{\alpha,r}$ is non-singular, problem in Eq. (\ref{eq: Transformed data - hypothesis testing problem})
is equivalent to:
\begin{equation}
\begin{cases}
\mathcal{H}_{0}\,:\, & ||\bm{B}||_{F}=0\,,\\
\mathcal{H}_{1}\,:\, & ||\bm{B}||_{F}>0\,,
\end{cases}
\end{equation}
which partitions the relevant-signal parameter space, say $\bm{\Theta}_{r}$,
as:
\begin{equation}
\bm{\Theta}_{r}=\underbrace{\{\bm{0}_{r\times M}\}}_{\bm{\Theta}_{r,0}}\cup\underbrace{\{\bm{B}\in\mathbb{C}^{r\times M}\,:\,\left\Vert \bm{B}\right\Vert _{F}>0\}}_{\bm{\Theta}_{r,1}}.\label{eq: formal_structure_hypothesis_testing}
\end{equation}
The canonical form in Eq. (\ref{eq: Transformed data - hypothesis testing problem})
will be exploited hereinafter in our analysis.

In the following, our analysis is carried out assuming that $(K-M)\geq N$.
Such condition is typically satisfied in practical adaptive detection
setups \cite{Kelly1989}.

\section{Maximal Invariant Statistic\label{sec: MIS}}

In what follows, we will search for decision rules sharing invariance
with respect to those parameters (namely the nuisance parameters,
$\bm{R}$, $\bm{B}_{t,1}$, and $\bm{B}_{t,0}$) which are irrelevant
for the specific decision problem. To this end, we resort to the so-called
``Principle of Invariance'' \cite{Lehmann2006}, whose main idea
consists in finding transformations that properly cluster data without
altering
\begin{itemize}
\item the formal structure of the hypothesis testing problem given by (\ref{eq: formal_structure_hypothesis_testing});
\item the Gaussian assumption for the received data matrix under each hypothesis;
\item the double-subspace structure containing the useful signal components. 
\end{itemize}
The following subsection is thus devoted to the definition of a suitable
group which fullfils the above requirements.

\subsection{Desired invariance properties \label{sub: Desired Invariance Properties}}

Let
\begin{equation}
\bm{V}_{c,1}\triangleq\begin{bmatrix}\bm{I}_{M}\\
\bm{0}_{(K-M)\times M}
\end{bmatrix},\quad\,\bm{V}_{c,2}\triangleq\begin{bmatrix}\bm{0}_{M\times(K-M)}\\
\bm{I}_{K-M}
\end{bmatrix},
\end{equation}
 and observe that $\bm{P}_{\bm{C}^{\dagger}}=(\bm{V}_{c,1}\bm{V}_{c,1}^{\dagger})$
and $\bm{P}_{\bm{C}^{\dagger}}^{\perp}=(\bm{V}_{c,2}\bm{V}_{c,2}^{\dagger})$.

Also, let us consider the sufficient statistic\footnote{Indeed, Fisher-Neyman factorization theorem ensures that deciding
from $\{\bm{Z}_{c},\bm{S}_{c}\}$ is tantamount to deciding from raw
data $\bm{Z}$ \cite{Muirhead2009}.} $\{\bm{Z}_{c},\bm{S}_{c}\}$, where the mentioned quantities are
defined as 
\begin{align}
\bm{Z}_{c}\triangleq\, & \left(\bm{Z}\bm{V}_{c,1}\right)\in\mathbb{C}^{N\times M}\\
\bm{Z}_{c,\perp}\triangleq\, & (\bm{Z}\,\bm{V}_{c,2})\in\mathbb{C}^{N\times(K-M)}\\
\bm{S}_{c}\triangleq\, & (\bm{Z}_{c,\perp}\bm{Z}_{c,\perp}^{\dagger})=(\bm{Z}\bm{P}_{\bm{C}^{\dagger}}^{\perp}\bm{Z}^{\dagger})\in\mathbb{C}^{N\times N}
\end{align}
Clearly, given the simplified structure of $\bm{C}$, $\bm{Z}_{c}$
(resp. $\bm{Z}_{c,\perp}$) is simply obtained by taking the first
$M$ (resp. the last $K-M$) columns of the transformed data matrix
$\bm{Z}$.

Now, denote by $\mathcal{GL}(N)$ the linear group of $N\times N$
non-singular matrices and introduce the following sets
\begin{eqnarray}
\mathcal{G} & \triangleq & \left\{ \bm{G}\triangleq\begin{bmatrix}\bm{G}_{11} & \bm{G}_{12} & \bm{G}_{13}\\
\bm{0}_{r\times t} & \bm{G}_{22} & \bm{G}_{23}\\
\bm{0}_{(N-J)\times t} & \bm{0}_{(N-J)\times r} & \bm{G}_{33}
\end{bmatrix}\in\mathcal{GL}(N)\right.\\
 &  & \left.\bm{G}:\bm{G}_{11}\in\mathcal{GL}(t),\,\bm{G}_{22}\in\mathcal{GL}(r),\,\bm{G}_{33}\in\mathcal{GL}(N-J)\right\} \nonumber 
\end{eqnarray}
\begin{equation}
\mathcal{F}\triangleq\left\{ \bm{F}\triangleq\begin{bmatrix}\bm{F}_{1}\\
\bm{0}_{r\times M}\\
\bm{0}_{(N-J)\times M}
\end{bmatrix}\in\mathbb{C}^{N\times M}\,:\,\bm{F}_{1}\in\mathbb{C}^{t\times M}\right\} 
\end{equation}
along with the composition operator ``$\circ$'', defined as:
\begin{equation}
(\bm{G}_{a},\bm{F}_{a})\circ(\bm{G}_{b},\bm{F}_{b})=(\bm{G}_{b}\bm{G}_{a},\bm{G}_{b}\bm{F}_{a}+\bm{F}_{b})\label{eq:  group composition operation}
\end{equation}
The sets and the composition operator are here represented compactly
as $\mathcal{L}\triangleq(\mathcal{G}\times\mathcal{F},\circ)$. Then,
it is not difficult to show that $\mathcal{L}$ constitutes a \emph{group},
since it satisfies the following elementary axioms:
\begin{itemize}
\item $\mathcal{L}$ is \emph{closed} with respect to the operation ``$\circ$'',
defined in Eq. (\ref{eq:  group composition operation});
\item $\forall(\bm{G}_{a},\bm{F}_{a})$, $(\bm{G}_{b},\bm{F}_{b})$, and
$(\bm{G}_{c},\bm{F}_{c})\in\mathcal{L}$: $[(\bm{G}_{a},\bm{F}_{a})\circ(\bm{G}_{b},\bm{F}_{b})]\circ(\bm{G}_{c},\bm{F}_{c})=(\bm{G}_{a},\bm{F}_{a})\circ[(\bm{G}_{b},\bm{F}_{b})\circ(\bm{G}_{c},\bm{F}_{c})]$
(\emph{Associative property});
\item there exists a unique $(\bm{G}_{I},\bm{F}_{I})\in\mathcal{L}$ such
that $\forall(\bm{G},\bm{F})\in\mathcal{L}$: $(\bm{G}_{I},\bm{F}_{I})\circ(\bm{G},\bm{F})=(\bm{G},\bm{F})\circ(\bm{G}_{I},\bm{F}_{I})=(\bm{G},\bm{F})$
(\emph{Existence of Identity element});
\item $\forall(\bm{G},\bm{F})\in\mathcal{L}$, there exists ($\bm{G}_{-1},\bm{F}_{-1})\in\mathcal{L}$
such that $(\bm{G}_{-1},\bm{F}_{-1})\circ(\bm{G},\bm{F})=(\bm{G},\bm{F})\circ(\bm{G}_{-1},\bm{F}_{-1})=(\bm{G}_{I},\bm{F}_{I})$
(\emph{Existence of Inverse element}).
\end{itemize}
Also, the aforementioned group leaves the hypothesis testing problem
in Eq. (\ref{eq: Transformed data - hypothesis testing problem})
invariant under the action $\ell$ defined by: 
\begin{equation}
\ell(\bm{Z}_{c},\bm{S}_{c})=\left(\bm{G}\bm{Z}_{c}+\bm{F},\bm{G}\,\bm{S}_{c}\,\bm{G}^{\dagger}\right)\quad\forall(\bm{G},\bm{F})\in\mathcal{L}\,.\label{eq: elementary action}
\end{equation}
The proof of the aforementioned statement is given in Appendix \ref{sec: Invariance wrt group L}.
Moreover, it is important to point out that $\mathcal{L}$ preserves
the family of distributions, and, at the same time, includes those
transformations which are relevant from a practical point of view,
as they allow claiming the CFAR property (with respect to $\bm{R}$
and $\bm{B}_{t,i}$) as a consequence of the invariance.

\subsection{Derivation of the MIS}

In Sec. \ref{sub: Desired Invariance Properties} we have identified
a group $\mathcal{L}$ which leaves unaltered the problem under investigation.
It is thus reasonable finding decision rules that are invariant under
$\mathcal{L}$. Toward this goal, the Principle of Invariance is invoked
because it allows to construct statistics that organize data into
distinguishable equivalence classes. Such functions of the data are
called \emph{Maximal Invariant Statistics }and, given the group of
transformations, every invariant test may be written as a function
of the maximal invariant \cite{Scharf1991}.

Before presenting the explicit expression of the MIS, we give the
following preliminary definitions based on the partitioning of matrices
$\bm{Z}_{c}$ and $\bm{S}_{c}$:
\begin{equation}
\bm{Z}_{c}=\begin{bmatrix}\bm{Z}_{1}\\
\bm{Z}_{2}\\
\bm{Z}_{3}
\end{bmatrix};\quad\bm{S}_{c}=\begin{bmatrix}\bm{S}_{11} & \bm{S}_{12} & \bm{S}_{13}\\
\bm{S}_{21} & \bm{S}_{22} & \bm{S}_{23}\\
\bm{S}_{31} & \bm{S}_{32} & \bm{S}_{33}
\end{bmatrix}\,.\label{eq: Z_c S_c (block definition)}
\end{equation}
where $\bm{Z}_{1}\in\mathbb{C}^{t\times M}$, $\bm{Z}_{2}\in\mathbb{C}^{r\times M}$,
and $\bm{Z}_{3}\in\mathbb{C}^{(N-J)\times M}$, respectively; $\bm{S}_{ij}$,
$(i,j)\in\{1,2,3\}\times\{1,2,3\}$, is a sub-matrix whose dimensions
can be obtained replacing $1$, $2$, and $3$ with $t$, $r$, and
$(N-J)$, respectively\footnote{Hereinafter, in the case $J=N$, the ``3-components'' are no longer
present in the partitioning.}. Furthermore, we also define the following partioning for $\bm{Z}_{c,\perp}$,
which will be used throughout the manuscript:
\begin{equation}
\bm{Z}_{c,\perp}=\begin{bmatrix}\bm{Z}_{\perp,1}^{T} & \bm{Z}_{\perp,2}^{T} & \bm{Z}_{\perp,3}^{T}\end{bmatrix}^{T}\label{eq: Z_c,perp partitioning}
\end{equation}
where $\bm{Z}_{\perp,1}\in\mathbb{C}^{t\times(K-M)}$, $\bm{Z}_{\perp,2}\in\mathbb{C}^{r\times(K-M)}$
and $\bm{Z}_{\perp,3}\in\mathbb{C}^{(N-J)\times(K-M)}$, respectively.
Observe that each sub-matrix of $\bm{S}_{c}$ in Eq. (\ref{eq: Z_c S_c (block definition)})
can be expressed in terms of Eq. (\ref{eq: Z_c,perp partitioning}),
that is, $\bm{S}_{ij}=(\bm{Z}_{\perp,i}\bm{Z}_{\perp,j}^{\dagger})$.
We are thus ready to present the proposition providing the expression
of a maximal invariant for the problem at hand. 
\begin{figure*}
\centering{}\includegraphics[width=0.6\paperwidth]{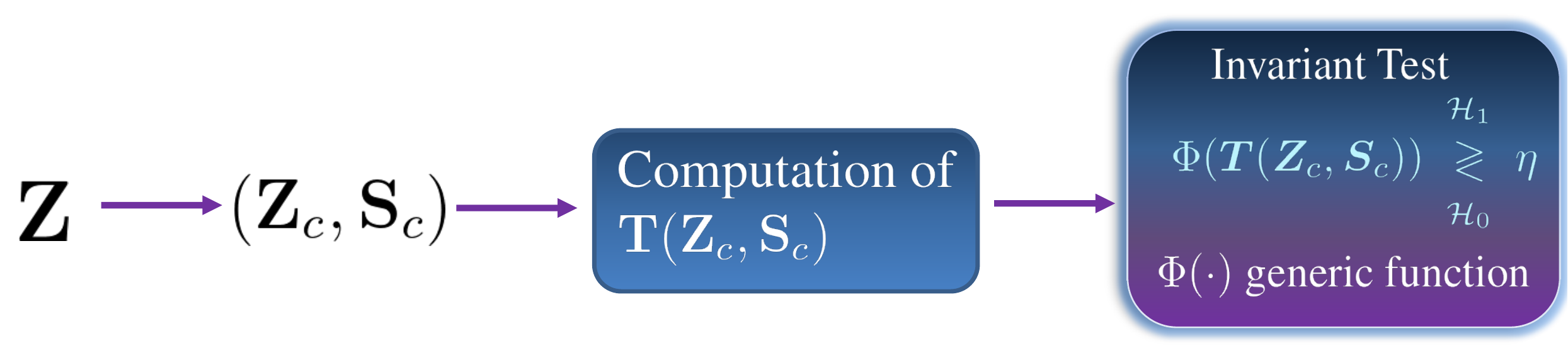}\protect\caption{Block diagram of processing leading to a generic invariant test.\label{fig: Block diagram Invariant Test}}
\end{figure*}

\begin{prop}
A MIS with respect to $\mathcal{L}$ for the problem in Eq. (\ref{eq: Transformed data - hypothesis testing problem})
is given by:\label{prop: Maximal Invariant Statistic}
\begin{gather}
\bm{T}(\bm{Z}_{c},\bm{S}_{c})=\begin{cases}
\begin{bmatrix}\bm{T}_{a}\triangleq\left\{ \bm{Z}_{2.3}^{\dagger}\,\bm{S}_{2.3}^{-1}\,\bm{Z}_{2.3}\right\} \\
\bm{T}_{b}\triangleq\left\{ \bm{Z}_{3}^{\dagger}\,\bm{S}_{33}^{-1}\bm{Z}_{3}\right\} 
\end{bmatrix} & J<N\\
\bm{Z}_{2}^{\dagger}\,\bm{S}_{22}^{-1}\bm{Z}_{2} & J=N
\end{cases}\label{eq: MIS_final}
\end{gather}
where $\bm{Z}_{2.3}\triangleq(\bm{Z}_{2}-\bm{S}_{23}\bm{S}_{33}^{-1}\bm{Z}_{3})$
and $\bm{S}_{2.3}\triangleq(\bm{S}_{22}-\bm{S}_{23}\bm{S}_{33}^{-1}\bm{S}_{32})$.\end{prop}
\begin{IEEEproof}
The proof is given in Appendix \ref{sec: Appendix _ MIS derivation}.
\end{IEEEproof}
Some important remarks are now in order.
\begin{itemize}
\item In the case $J<N$, the MIS is given by a pair of matrices (namely
$\bm{T}_{a}$ and $\bm{T}_{b}$) where the second component ($\bm{T}_{b}$)
represents an \emph{ancillary part, }that is, its distribution does
not depend on the hypothesis in force;
\item In the case $J<N$, the matrices $\bm{T}_{a}\in\mathbb{C}^{M\times M}$
and $\bm{T}_{b}\in\mathbb{C}^{M\times M}$ have rank equal to $\min\{M,r\}$
and $\min\{M,N-J\}$, respectively;
\item It is of certain interest comparing the general expression in Eq.
(\ref{eq: MIS_final}) with the MIS instances obtained in \cite{Raghavan1996,A.DeMaio2014,Bose1995,Conte2003}
for specific adaptive detection scenarios. Accordingly, Sec. \ref{sec: MIS (special instances)}
will be devoted to comparisons and exhaustive discussion of the specialized
forms in some relevant scenarios;
\item Finally, exploiting \cite[Thm. 6.2.1]{Lehmann2006}, every invariant
test may be written as a function of Eq. (\ref{eq: MIS_final}) (see.
Fig. \ref{fig: Block diagram Invariant Test} for a schematic representation).
Therefore, it naturally follows that every CFAR test can be expressed
in terms of the MIS. Part II of this study will be devoted to the
design of theoretically-founded detectors whose CFARness will be proved
by showing their dependence on the data solely through the obtained
MIS.
\end{itemize}

\section{Statistical Characterization of the MIS\label{sec: Stat Char MIS}}

In this section, we provide the statistical characterization of the
MIS for the case $J<N$ and then, we will give a corollary referring
to $J=N$. To this end, we show that the MIS can be written as a function
of whitened random vectors and matrices and then we find a suitable
stochastic representation by means of a one-to-one transformation. 

First, we consider the following transformation $(\bm{G}^{\circ},\bm{F}^{\circ})\in\mathcal{L}$,
which leads to:
\begin{equation}
\bm{Z}_{c}^{\circ}=\bm{G}^{\circ}\bm{Z}_{c}+\bm{F}^{\circ}=\begin{bmatrix}\bm{Z}_{1}^{\circ T} & \bm{Z}_{2}^{\circ T} & \bm{Z}_{3}^{\circ T}\end{bmatrix}^{T}
\end{equation}
and
\begin{equation}
\bm{S}_{c}^{\circ}=\bm{G}^{\circ}\bm{S}_{c}\,\bm{G}^{\circ\dagger}=\begin{bmatrix}\bm{S}_{11}^{\circ} & \bm{S}_{12}^{\circ} & \bm{S}_{13}^{\circ}\\
\bm{S}_{21}^{\circ} & \bm{S}_{22}^{\circ} & \bm{S}_{23}^{\circ}\\
\bm{S}_{31}^{\circ} & \bm{S}_{32}^{\circ} & \bm{S}_{33}^{\circ}
\end{bmatrix}
\end{equation}
where $\bm{Z}_{i}^{\circ}$ and $\bm{S}_{\ell m}^{\circ}$ ($i,\ell,m\in\{1,2,3\}$)
are similarly defined as in Eq. (\ref{eq: Z_c S_c (block definition)})
and the pair $(\bm{G}^{\circ},\bm{F}^{\circ})$ is suitably defined
as:
\begin{align}
\bm{G}^{\circ}\triangleq\, & \begin{bmatrix}\bm{G}_{11}^{\circ} & \bm{G}_{12}^{\circ} & \bm{G}_{13}^{\circ}\\
\bm{0}_{r\times t} & \bm{R}_{2.3}^{-1/2} & -\bm{R}_{2.3}^{-1/2}\bm{R}_{23}\,\bm{R}_{33}^{-1}\\
\bm{0}_{(N-J)\times t} & \bm{0}_{(N-J)\times r} & \bm{R}_{33}^{-1/2}
\end{bmatrix};\\
\bm{F}^{\circ}\triangleq\, & \begin{bmatrix}\bm{F}_{1}^{\circ}\\
\bm{0}_{r\times M}\\
\bm{0}_{(N-J)\times M}
\end{bmatrix};
\end{align}
where $\bm{G}_{i,j}^{\circ}$, $i=1$, $j\in\{1,2,3\}$, and $\bm{F}_{1}^{\circ}$
are generic matrices of proper dimensions, while $\bm{R}_{22}\in\mathbb{C}^{r\times r}$,
$\bm{R}_{23}\in\mathbb{C}^{r\times(N-J)}$, and $\bm{R}_{33}\in\mathbb{C}^{(N-J)\times(N-J)}$
are obtained partitioning the true covariance matrix $\bm{R}$ in
the same way as done for $\bm{S}$ in Eq.~(\ref{eq: Z_c S_c (block definition)}).
Finally, we have defined $\bm{R}_{2.3}\triangleq\bm{R}_{22}-\bm{R}_{23}\,\bm{R}_{33}^{-1}\,\bm{R}_{23}^{\dagger}$. 

Hereinafter, we will study MIS statistical characterization after
the trasformation $(\bm{G}^{\circ},\bm{F}^{\circ})$. This will simplify
the subsequent analysis and does not affect the obtained results since
the MIS is (by definition) invariant with respect to every trasformation
belonging to $\mathcal{L}$. 

Now observe that, under $\mathcal{H}_{i}$, $i\in\{0,1\}$, it holds:
\begin{gather}
\begin{bmatrix}\bm{Z}_{2}^{\circ}\\
\bm{Z}_{3}^{\circ}
\end{bmatrix}|\mathcal{H}_{i}=\bm{G}_{3}^{\circ}\begin{bmatrix}\bm{Z}_{2}\\
\bm{Z}_{3}
\end{bmatrix}|\mathcal{H}_{i}\nonumber \\
\sim\mathcal{CN}_{(N-t)\times M}\left(i\,\bm{G}_{3}^{\circ}\,\begin{bmatrix}\bm{I}_{r}\\
\bm{0}_{(N-J)\times r}
\end{bmatrix}\bm{B},\,\bm{I}_{M},\,\bm{I}_{N-t}\right)\\
\begin{bmatrix}\bm{S}_{22}^{\circ} & \bm{S}_{23}^{\circ}\\
\bm{S}_{32}^{\circ} & \bm{S}_{33}^{\circ}
\end{bmatrix}=\left\{ \bm{G}_{3}^{\circ}\,\begin{bmatrix}\bm{S}_{22} & \bm{S}_{23}\\
\bm{S}_{32} & \bm{S}_{33}
\end{bmatrix}\,(\bm{G}_{3}^{\circ})^{\dagger}\right\} \nonumber \\
\sim\mathcal{C}\mathcal{W}_{N-t}(K-M,\,\bm{I}_{N-t})
\end{gather}
where we have defined $\bm{G}_{3}^{\circ}\in\mathbb{C}^{(N-t)\times(N-t)}$
as
\begin{equation}
\bm{G}_{3}^{\circ}\triangleq\begin{bmatrix}\bm{R}_{2.3}^{-1/2} & -\bm{R}_{2.3}^{-1/2}\,\bm{R}_{23}\,\bm{R}_{33}^{-1}\\
\bm{0}_{(N-J)\times r} & \bm{R}_{33}^{-1/2}
\end{bmatrix}\,.
\end{equation}
Thus, exploiting the invariance property, we can equivalently rewrite
the two components of $\bm{T}(\bm{Z}_{c},\bm{S}_{c})$ (cf. Eq. (\ref{eq: MIS_final}))
in terms of the \emph{whitened} quantities
\begin{align}
\bm{T}_{a} & =(\bm{Z}_{2.3}^{\circ})^{\dagger}\,(\bm{S}_{2.3}^{\circ})^{-1}\,\bm{Z}_{2.3}^{\circ},\\
\bm{T}_{b} & =(\bm{Z}_{3}^{\circ})^{\dagger}\,(\bm{S}_{33}^{\circ})^{-1}\,\bm{Z}_{3}^{\circ},
\end{align}
where $\bm{Z}_{2.3}^{\circ}\triangleq\bm{Z}_{2}^{\circ}-\bm{S}_{23}^{\circ}\,(\bm{S}_{33}^{\circ})^{-1}\,\bm{Z}_{3}^{\circ}$
and $\bm{S}_{2.3}^{\circ}\triangleq\bm{S}_{22}^{\circ}-\bm{S}_{23}^{\circ}(\bm{S}_{33}^{\circ})^{-1}\bm{S}_{32}^{\circ}$,
respectively. Let us focus on $\bm{Z}_{2.3}^{\circ}$ and rewrite
\begin{eqnarray}
\bm{Z}_{2}^{\circ} & = & \begin{bmatrix}\bm{z}_{2,1}^{\circ} & \cdots & \bm{z}_{2,M}^{\circ}\end{bmatrix}\\
\bm{S}_{23}^{\circ}\,(\bm{S}_{33}^{\circ})^{-1}\,\bm{Z}_{3}^{\circ} & = & \sum_{k=1}^{(K-M)}\bm{r}_{2,k}\,\bm{r}_{3,k}^{\dagger}\,(\bm{S}_{33}^{\circ})^{-1}\,\bm{Z}_{3}^{\circ}
\end{eqnarray}
where $\bm{r}_{j,k}$ generically denotes the $k$-th column of $\bm{Z}_{\perp,j}$.
Given the aforementioned definitions, we obtain the explicit form
given in Eq.~(\ref{eq: three_com_MIS_dist}) (at the top of next
page) for $\bm{Z}_{2.3}^{\circ}$ in terms of its columns $\bm{q}_{\ell}$,
$\ell=1,\ldots M$, where we have further defined $\bm{a}_{\ell}\triangleq\left\{ (\bm{S}_{33}^{\circ})^{-1}\,\bm{z}_{3,\ell}^{\circ}\right\} $
and $\bm{z}_{3,\ell}$ similarly represents the $\ell$-th column
of $\bm{Z}_{3}^{\circ}$, that is $\bm{Z}_{3}^{\circ}=\begin{bmatrix}\bm{z}_{3,1}^{\circ} & \cdots & \bm{z}_{3,M}^{\circ}\end{bmatrix}$.
\begin{figure*}
\begin{equation}
\bm{Z}_{2.3}^{\circ}=\begin{bmatrix}\bm{q}_{1} & \cdots & \bm{q}_{M}\end{bmatrix}=\begin{bmatrix}\left(\bm{z}_{2,1}^{\circ}-\sum_{k=1}^{K-M}\bm{r}_{2,k}\bm{r}_{3,k}^{\dagger}\bm{a}_{1}\right) & \cdots & \left(\bm{z}_{2,M}^{\circ}-\sum_{k=1}^{K-M}\bm{r}_{2,k}\bm{r}_{3,k}^{\dagger}\bm{a}_{M}\right)\end{bmatrix}\label{eq: three_com_MIS_dist}
\end{equation}
\hrulefill 
\vspace*{0pt}
\end{figure*}

First, we observe that $\bm{r}_{2,k}\sim\mathcal{CN}_{r}(\bm{0}_{r},\bm{I}_{r})$
and $\bm{r}_{3,k}\sim\mathcal{C}\mathcal{N}_{N-J}(\bm{0}_{N-J},\bm{I}_{N-J})$;
also it is apparent that these vectors are all mutually independent.
Before proceeding, we define the short-hand notation ``$\text{\#3}$''
to denote the conditioning with respect to all the terms with subscript
``$3$''. Then, it can be shown that $\bm{q}_{\ell}|(\#3,\mathcal{H}_{0})$
is Gaussian distributed (recall that $\bm{z}_{2,\ell}|\mathcal{H}_{0}\sim\mathcal{CN}_{r}(\bm{0}_{r},\bm{I}_{r})$)
with mean vector $\bm{0}_{r}$ and covariance:
\begin{gather}
\mathbb{E}\left\{ (\bm{z}_{2,\ell}^{\circ}-\sum_{k=1}^{K-M}\bm{r}_{2,k}\bm{r}_{3,k}^{\dagger}\bm{a}_{\ell})(\bm{z}_{2,\ell}^{\circ}-\sum_{k=1}^{K-M}\bm{r}_{2,k}\bm{r}_{3,k}^{\dagger}\bm{a}_{\ell})^{\dagger}\right\} =\nonumber \\
\bm{I}_{r}\left(1+\bm{z}_{3,\ell}^{\dagger}\,(\bm{S}_{33}^{\circ})^{-1}\bm{z}_{3,\ell}\right)
\end{gather}
 Similarly, the cross-covariance between $\bm{q}_{\ell}|(\#3,\mathcal{H}_{0})$
and $\bm{q}_{m}|(\#3,\mathcal{H}_{0})$ is given by:
\begin{gather}
\mathbb{E}\left\{ (\bm{z}_{2,m}^{\circ}-\sum_{k=1}^{K-M}\bm{r}_{2,k}\bm{r}_{3,k}^{\dagger}\bm{a}_{m})(\bm{z}_{2,\ell}^{\circ}-\sum_{k=1}^{K-M}\bm{r}_{2,k}\bm{r}_{3,k}^{\dagger}\bm{a}_{\ell})^{\dagger}\right\} =\nonumber \\
\bm{I}_{r}\left(\bm{z}_{3,m}^{\dagger}\,(\bm{S}_{33}^{\circ})^{-1}\bm{z}_{3,\ell}\right)
\end{gather}
Therefore, in view the aforementioned results, it follows that $\bm{\zeta}_{2.3}\triangleq\mathrm{vec}(\bm{Z}_{2.3}^{\circ})$
is conditionally distributed as:
\begin{gather}
\bm{\zeta}_{2.3}|(\#3,\mathcal{H}_{0})\sim\mathcal{CN}_{rM}\left(\bm{0}_{rM},(\bm{I}_{M}+(\bm{Z}_{3}^{\circ})^{\dagger}(\bm{S}_{33}^{\circ})^{-1}\bm{Z}_{3}^{\circ})\otimes\bm{I}_{r}\right)
\end{gather}
Then, we whiten $\bm{\zeta}_{2.3}$, that is, we define:
\begin{gather}
\bm{x}\triangleq\left[(\bm{I}_{M}+\bm{Z}_{3}^{\circ\dagger}(\bm{S}_{33}^{\circ})^{-1}\bm{Z}_{3}^{\circ})\otimes\bm{I}_{r}\right]^{-1/2}\bm{\xi}_{2.3}
\end{gather}
which evidently gives $\bm{x}|(\#3,\mathcal{H}_{0})\sim\mathcal{CN}_{rM}(\bm{0}_{rM},\bm{I}_{rM})$.
Also, we observe that:
\begin{gather}
\left[\left(\bm{I}_{M}+\bm{Z}_{3}^{\circ\dagger}(\bm{S}_{33}^{\circ})^{-1}\bm{Z}_{3}^{\circ}\right)\otimes\bm{I}_{r}\right]^{-1/2}=\\
\left[\left(\bm{I}_{M}+\bm{Z}_{3}^{\circ\dagger}(\bm{S}_{33}^{\circ})^{-1}\bm{Z}_{3}^{\circ}\right)^{-1/2}\right]\otimes\bm{I}_{r}\nonumber 
\end{gather}
which readily follows from the distributive property of Kronecker
product. As a consequence, we have that $\bm{x}=\mathrm{vec}(\bm{X})$,
where we have defined
\begin{equation}
\bm{X}\triangleq(\bm{Z}_{2.3}^{\circ}\,\bm{K}_{s3})\,,
\end{equation}
and $\bm{K}_{s3}\triangleq\left[(\bm{I}_{M}+\bm{Z}_{3}^{\circ\dagger}\,(\bm{S}_{33}^{\circ})^{-1}\bm{Z}_{3}^{\circ})^{-1/2}\right]^{T}$.
Also, it is straightforward to show that $\bm{X}|(\mathcal{H}_{0},\#3)=\bm{X}|\mathcal{H}_{0}\sim\mathcal{CN}_{r\times M}(\bm{0}_{r\times M},\,\bm{I}_{M},\,\bm{I}_{r})$
(i.e., it does not depend on the components with subscript ``3'').
We then consider a one-to-one transformation of $\bm{T}(\bm{Z}_{c}^{\circ},\,\bm{S}_{c}^{\circ})$,
defined as:
\begin{align}
\bm{T}_{1}(\bm{Z}_{c}^{\circ},\,\bm{S}_{c}^{\circ}) & \triangleq\begin{bmatrix}\bm{K}_{s3}^{\dagger}\,(\bm{Z}_{2.3}^{\circ})^{\dagger}\,(\bm{S}_{2.3}^{\circ})^{-1}\,\bm{Z}_{2.3}^{\circ}\,\bm{K}_{s3}\\
(\bm{Z}_{3}^{\circ})^{\dagger}\,(\bm{S}_{33}^{\circ})^{-1}\,\bm{Z}_{3}^{\circ}
\end{bmatrix}\\
 & =\begin{bmatrix}\bm{X}^{\dagger}\,(\bm{S}_{2.3}^{\circ})^{-1}\,\bm{X}\\
(\bm{Z}_{3}^{\circ})^{\dagger}\,(\bm{S}_{33}^{\circ})^{-1}\,\bm{Z}_{3}^{\circ}
\end{bmatrix}\triangleq\begin{bmatrix}\bm{T}_{1,a}\\
\bm{T}_{1,b}
\end{bmatrix}
\end{align}
It is clear that, since $\bm{T}_{1}(\cdot)$ is a one-to-one transformation
of the MIS, it is a MIS itself \cite{Lehmann2006}. Therefore, without
loss of generality, we will concentrate on the statistical characterization
of $\bm{T}_{1}(\cdot)$.

We start by recalling that $\bm{S}_{2.3}^{\circ}$ is independent
of $\{\bm{S}_{22}^{\circ},\bm{S}_{33}^{\circ}\}$ \cite[Thm. A.11]{Bandiera2009}.
Also, we notice that $\bm{S}_{33}^{\circ}\sim\mathcal{CW}_{N-J}(K-M,\,\bm{I}_{N-J})$
and $\bm{S}_{2.3}^{\circ}\sim\mathcal{CW}_{r}((K-M)-(N-J),\,\bm{I}_{r})$.
These results hold under both the hypotheses. Furthermore, conditioned
on $\mathcal{H}_{0}$, $\bm{X}$ is independent on $\bm{T}_{1,b}$
(as it is independent on terms with subscript ``$3$''). 

Therefore, it follows that conditioned on $\mathcal{H}_{0}$, $\bm{T}_{1,a}$
and $\bm{T}_{1,b}$ are \emph{statistically independent matrices},
which means that the joint pdf can be written as $f_{0}(\bm{T}_{1,a},\bm{T}_{1,b})=f_{0}(\bm{T}_{1,a})\,f(\bm{T}_{1,b})$
(as $\bm{T}_{1,b}$ denotes the ancillary part of the MIS and thus
its pdf is independent on the specific hypothesis). Finally, it is
worth noticing that in the case $M\leq r$ we obtain the explicit
pdf of $\bm{T}_{1,a}|\mathcal{H}_{0}\sim\mathcal{CF}_{M}(\bm{0}_{M\times M},\,r,\,(K-M)-(N-J))$
and, for $M\leq(N-J)$, $\bm{T}_{1,b}\sim\mathcal{CF}_{M}(\bm{0}_{M\times M},\,N-J,\,(K-M)-(N-J))$,
following \cite{James1964}.

On the other hand, when $\mathcal{H}_{1}$ holds true, it is not difficult
to show that $\bm{Z}_{2.3}^{\circ}|(\mathcal{H}_{1},\#3)\sim\mathcal{CN}_{r\times M}(\,\bm{R}_{2.3}^{-1/2}\,\bm{B},\,(\bm{I}_{M}+(\bm{Z}_{3}^{\circ})^{\dagger}\,(\bm{S}_{33}^{\circ})^{-1}\bm{Z}_{3}^{\circ}),\,\bm{I}_{r})$
and consequently $\bm{X}|(\mathcal{H}_{1},\#3)\sim\mathcal{CN}_{r\times M}(\,\bm{R}_{2.3}^{-1/2}\,\bm{B}\,\bm{K}_{s3},\,\bm{I}_{M},\,\bm{I}_{r})$.
A direct inspection of the last result reveals that 
\begin{gather}
\bm{X}|(\#3,\mathcal{H}_{1})=\bm{X}|(\bm{T}_{1,b},\mathcal{H}_{1})\sim\nonumber \\
\mathcal{CN}_{r\times M}\left(\bm{R}_{2.3}^{-1/2}\,\bm{B}\,\left((\bm{I}_{M}+\bm{T}_{1,b})^{-1/2}\right)^{T},\,\bm{I}_{M},\,\bm{I}_{r}\right)
\end{gather}
which underlines that $\bm{T}_{1,a}$ and $\bm{T}_{1,b}$ are statistically
dependent under $\mathcal{H}_{1}$, thus leading to $f_{1}(\bm{T}_{1,a},\bm{T}_{1,b})=f_{1}(\bm{T}_{1,a}|\bm{T}_{1,b})\,f(\bm{T}_{b})$.
Again, in the specific case $M\leq r$ it holds $\bm{T}_{1.a}|(\mathcal{H}_{1},\bm{T}_{1,b})\sim\mathcal{CF}_{M}(\bm{\Omega},\,r,\,(K-M)-(N-J))$,
where we have denoted $\bm{\Omega}\triangleq(\bm{K}_{s3}^{\dagger}\,\bm{B}^{\dagger}\,\bm{R}_{2.3}^{-1}\,\bm{B}\,\bm{K}_{s3})$,
following \cite{James1964}. 

Finally, we conclude the section with a discussion on the induced
maximal invariant in the parameter space \cite{Lehmann2006}. The
induced maximal invariant represents the reduced set of unknown parameters
on which the hypothesis testing in the invariant domain depends. It
is not difficult to show that for I-GMANOVA model this equals to $\bm{T}_{\mathrm{p}}\triangleq\bm{B}^{\dagger}\,\bm{R}_{2.3}^{-1}\,\bm{B}\in\mathbb{C}^{M\times M}$.
In addition, the induced maximal invariant is not full rank in the
general case, with corresponding rank being equal to $\min\{r,\,M\}$.
It is worth remarking that such result applies in general, that is,
the distribution of the MIS will depend on the parameter space only
through $\bm{T}_{\mathrm{p}}$, following classic results from \cite{Lehmann2006}.

\section{MIS in special cases\label{sec: MIS (special instances)}}

\subsection{Adaptive detection of a point-like target}

In the present case we start from general formulation in Eq.~(\ref{eq: Hypothesis testing formulation})
and assume that: ($i$) $t=0$ (i.e., there is no interference); $(ii)$
$r=1$ (thus $J=1$) i.e., the matrix $\widetilde{\bm{A}}_{r}$ collapses
to $\tilde{\bm{a}}_{r}\in\mathbb{C}^{N\times1}$; ($iii$) $M=1$,
i.e. the matrix $\widetilde{\bm{B}}_{r}$ collapses to a scalar $\widetilde{b}_{r}\in\mathbb{C}$
and $(iv)$ $\tilde{\bm{c}}\triangleq\begin{bmatrix}1 & 0 & \cdots & 0\end{bmatrix}\in\mathbb{C}^{1\times K}$
(i.e. a row vector). Such case has been dealt in \cite{Bose1995}.
Therefore, the hypothesis testing in canonical form is given by: 
\begin{equation}
\begin{cases}
\mathcal{H}_{0}: & \bm{Z}=\bm{N}\\
\mathcal{H}_{1}: & \bm{Z}=\bm{a}\,b\,\bm{c}+\bm{N}
\end{cases}
\end{equation}
where $\bm{a}=\begin{bmatrix}1 & 0 & \cdots & 0\end{bmatrix}^{T}\in\mathbb{C}^{N\times1}$
and $\bm{c}=\widetilde{\bm{c}}$. By looking at the general MIS statistic
expression in Eq. (\ref{eq: MIS_final}), it is not difficult to show
that the present problem admits the following simplified partioning:
\begin{equation}
\bm{z}_{c}=\begin{bmatrix}z_{2}\\
\bm{z}_{3}
\end{bmatrix},\quad\bm{Z}_{c,\perp}=\begin{bmatrix}\bm{z}_{\perp,2}\\
\bm{Z}_{\perp,3}
\end{bmatrix},\quad\bm{S}_{c}=\begin{bmatrix}s_{22} & \bm{s}_{23}\\
\bm{s}_{32} & \bm{S}_{33}
\end{bmatrix},
\end{equation}
where $z_{2}\in\mathbb{C}$, $\bm{z}_{3}\in\mathbb{C}^{(N-1)\times1}$,
$\bm{z}_{\perp,2}\in\mathbb{C}^{1\times(K-1)}$ (i.e., a row vector),
$\bm{Z}_{\perp,3}\in\mathbb{C}^{(N-1)\times(K-1)}$, $s_{22}\in\mathbb{C}$,
$\bm{s}_{23}\in\mathbb{C}^{1\times(N-1)}$ (i.e. a row vector), $\bm{s}_{32}\in\mathbb{C}^{(N-1)\times1}$
and $\bm{S}_{33}\in\mathbb{C}^{(N-1)\times(N-1)}$, respectively.
Exploiting the above partioning, gives the simplified expressions:
\begin{eqnarray}
s_{2.3} & = & (s_{22}-\bm{s}_{23}\,\bm{S}_{33}^{-1}\,\bm{s}_{32})\\
z_{2.3} & = & \left(z_{2}-\bm{s}_{23}\,\bm{S}_{33}^{-1}\,\bm{z}_{3}\right)
\end{eqnarray}
which are both scalar valued. Therefore, it is not difficult to show
that the two components of the MIS are both \emph{scalar valued} and
equal to:
\begin{align}
t_{a} & =\frac{|z_{2}-\bm{s}_{23}\,\bm{S}_{33}^{-1}\,\bm{z}_{3}|^{2}}{\left[s_{22}-\bm{s}_{23}\,\bm{S}_{33}^{-1}\,\bm{s}_{32}\right]}\\
t_{b} & =\bm{z}_{3}^{\dagger}\,\bm{S}_{33}^{-1}\,\bm{z}_{3}
\end{align}
which can be rewritten in the more familiar form:
\begin{align}
t_{a} & =\frac{\left|z_{2}-\bm{z}_{\perp,2}\bm{Z}_{\perp,3}^{\dagger}\,(\bm{Z}_{\perp,3}\bm{Z}_{\perp,3}^{\dagger})^{-1}\bm{z}_{3}\right|^{2}}{\bm{z}_{\perp,2}\left[\bm{I}_{K-1}-\bm{Z}_{\perp,3}^{\dagger}(\bm{Z}_{\perp,3}\bm{Z}_{\perp,3}^{\dagger})^{-1}\bm{Z}_{\perp,3}\right]\bm{z}_{\perp,2}^{\dagger}}\\
t_{b} & =\bm{z}_{3}^{\dagger}(\bm{Z}_{\perp,3}\bm{Z}_{\perp,3}^{\dagger})^{-1}\bm{z}_{3}
\end{align}
which are those obtained in \cite{Bose1995}. 

Finally, it is not difficult to show that in such case the (scalar-valued)
induced maximal invariant is $t_{\mathrm{p}}=(|b|^{2}/r_{2.3})$,
where $r_{2.3}\triangleq(r_{22}-\bm{r}_{23}\,\bm{R}_{33}^{-1}\,\bm{r}_{32})$,
where we exploited similar simplified partitioning for $\bm{R}$ as
for $\bm{S}_{c}$. The induced maximal invariant clearly coincides
with the Signal-to-Noise plus Interference Ratio (SINR).

\subsection{Adaptive vector subspace detection}

In the present case we start from the general formulation in Eq.~(\ref{eq: Hypothesis testing formulation})
and assume that: ($i$) $t=0$ (i.e. there is no interference, thus
$J=r$); ($ii$) $M=1$, i.e. the matrix $\widetilde{\bm{B}}_{r}$
collapses to a vector $\widetilde{\bm{b}}_{r}\in\mathbb{C}^{J\times1}$
and $(iii)$ $\tilde{\bm{c}}\triangleq\begin{bmatrix}1 & 0 & \cdots & 0\end{bmatrix}\in\mathbb{C}^{1\times K}$
(i.e., a row vector). Such case has been dealt in \cite{Scharf1994,Raghavan1996}.
Therefore, the hypothesis testing in canonical form is given by: 
\begin{equation}
\begin{cases}
\mathcal{H}_{0}: & \bm{Z}=\bm{N}\\
\mathcal{H}_{1}: & \bm{Z}=\bm{A}\,\bm{b}\,\bm{c}+\bm{N}
\end{cases}
\end{equation}
where $\bm{A}=\begin{bmatrix}\bm{I}_{r} & \bm{0}_{r\times(N-r)}\end{bmatrix}^{T}\in\mathbb{C}^{N\times r}$
and $\bm{c}=\widetilde{\bm{c}}$. By looking at the general MIS statistic
expression in Eq. (\ref{eq: MIS_final}) it is not difficult to show
that the present problem admits the following simplified partioning:
\begin{equation}
\bm{z}_{c}=\begin{bmatrix}\bm{z}_{2}\\
\bm{z}_{3}
\end{bmatrix},\quad\bm{Z}_{c,\perp}=\begin{bmatrix}\bm{Z}_{\perp,2}\\
\bm{Z}_{\perp,3}
\end{bmatrix},\quad\bm{S}_{c}=\begin{bmatrix}\bm{S}_{22} & \bm{S}_{23}\\
\bm{S}_{32} & \bm{S}_{33}
\end{bmatrix},
\end{equation}
where $\bm{z}_{2}\in\mathbb{C}^{J\times1}$, $\bm{z}_{3}\in\mathbb{C}^{(N-J)\times1}$,
$\bm{Z}_{\perp,2}\in\mathbb{C}^{J\times(K-1)}$, $\bm{Z}_{\perp,3}\in\mathbb{C}^{(N-J)\times(K-1)}$,
$\bm{S}_{22}\in\mathbb{C}^{J\times J}$, $\bm{S}_{23}\in\mathbb{C}^{J\times(N-J)}$,
$\bm{S}_{32}\in\mathbb{C}^{(N-J)\times J}$ and $\bm{S}_{33}\in\mathbb{C}^{(N-J)\times(N-J)}$,
respectively.

Exploiting the above partitioning, gives $\bm{S}_{2.3}\in\mathbb{C}^{J\times J}$
and the simplified expression:
\begin{equation}
\bm{z}_{2.3}=(\bm{z}_{2}-\bm{S}_{23}\,\bm{S}_{33}^{-1}\,\bm{z}_{3})\in\mathbb{C}^{J\times1}
\end{equation}
Since $\bm{z}_{2.3}$ and $\bm{z}_{3}$ are both column vectors, it
is not difficult to show that the two components of the MIS are both
\emph{scalar valued} and equal to:
\begin{gather}
t_{a}=\bm{z}_{2.3}^{\dagger}\,\bm{S}_{2.3}^{-1}\,\bm{z}_{2.3},\qquad t_{b}=\bm{z}_{3}^{\dagger}\,\bm{S}_{33}^{-1}\,\bm{z}_{3},\label{eq: MIS - vector subspace detection}
\end{gather}
which is the classic result obtained in \cite{Raghavan1996}.

Finally, the (scalar-valued) induced maximal invariant equals $t_{\mathrm{p}}=\bm{b}^{\dagger}\bm{R}_{2.3}^{-1}\bm{b}$
which is the result obtained in \cite{Raghavan1996}, being equal
to the SINR.

\subsection{Adaptive vector subspace detection with\protect \\
 structured interference}

In the present case we start from general formulation in Eq.~(\ref{eq: Hypothesis testing formulation})
and assume that: ($i$) $M=1$, i.e., the matrices $\widetilde{\bm{B}}_{r}$
and $\widetilde{\bm{B}}_{t}$ collapse to the vectors $\widetilde{\bm{b}}_{r}\in\mathbb{C}^{r\times1}$
and $\widetilde{\bm{b}}_{t}\in\mathbb{C}^{t\times1}$, respectively;
$(ii)$ $\tilde{\bm{c}}\triangleq\begin{bmatrix}1 & 0 & \cdots & 0\end{bmatrix}\in\mathbb{C}^{1\times K}$
(i.e. a row vector). Such case has been dealt in \cite{A.DeMaio2014}.
Given the aforementioned assumptions, the problem in canonical form
is given as: 
\begin{equation}
\begin{cases}
\mathcal{H}_{0}: & \bm{Z}=\bm{A}\,\begin{bmatrix}\bm{b}_{t,0}^{T} & \bm{0}_{r}^{T}\end{bmatrix}^{T}\,\bm{c}+\bm{N}\\
\mathcal{H}_{1}: & \bm{Z}=\bm{A}\,\begin{bmatrix}\bm{b}_{t,1}^{T} & \bm{b}^{T}\end{bmatrix}^{T}\,\bm{c}+\bm{N}
\end{cases}
\end{equation}
where $\bm{b}_{t,i}\in\mathbb{C}^{t\times1}$, $\bm{b}\in\mathbb{C}^{r\times1}$
and $\bm{c}=\widetilde{\bm{c}}$. By looking at the general MIS statistic
expression in Eq. (\ref{eq: MIS_final}) it is not difficult to show
that the present problem admits the following simplified expression:
\begin{equation}
\bm{z}_{c}^{T}=\begin{bmatrix}\bm{z}_{1}^{T} & \bm{z}_{2}^{T} & \bm{z}_{3}^{T}\end{bmatrix}^{T}
\end{equation}
where $\bm{z}_{1}\in\mathbb{C}^{t\times1}$, $\bm{z}_{2}\in\mathbb{C}^{r\times1}$
and $\bm{z}_{3}\in\mathbb{C}^{(N-J)\times1}$, respectively. Therefore,
it is readily shown that $\bm{S}_{2.3}\in\mathbb{C}^{r\times r}$
and $\bm{z}_{2.3}=(\bm{z}_{2}-\bm{S}_{23}\,\bm{S}_{33}^{-1}\,\bm{z}_{3})\in\mathbb{C}^{r\times1}$.
Since $\bm{z}_{2.3}$ and $\bm{z}_{3}$ are both column vectors, it
is not difficult to show that both components of the MIS are scalar-valued
(similarly to the ``no-interference'' case) and equal to:
\begin{gather}
t_{a}=\bm{z}_{2.3}^{\dagger}\,\bm{S}_{2.3}^{-1}\,\bm{z}_{2.3}\,;\qquad t_{b}=\bm{z}_{3}^{\dagger}\,\bm{S}_{33}^{-1}\,\bm{z}_{3}\,;\label{eq: MIS_vector subspace det plus Interference}
\end{gather}
which can be recognized as the result obtained in \cite{A.DeMaio2014}.
It is worth noticing that this result seems identical to that obtained
in the previous sub-section (i.e., the interference-free case). However,
we observe that, as opposed to the expression in (\ref{eq: MIS - vector subspace detection}),
definition of constituents of MIS in Eq. (\ref{eq: MIS_vector subspace det plus Interference})
is obtained by discarding terms with subscript ``1''. In other terms,
Eq. (\ref{eq: MIS_vector subspace det plus Interference}) is analogous
to (\ref{eq: MIS - vector subspace detection}) \emph{only} after
projection in the complementary subspace of the interference.

Finally, the (scalar-valued) induced maximal invariant is $t_{\mathrm{p}}=\bm{b}^{\dagger}\bm{R}_{2.3}^{-1}\,\bm{b}$,
which coincides with the result obtained in \cite{A.DeMaio2014},
being equal to the SINR in the complementary interference subspace.

\subsection{Multidimensional signals\label{sub: MIS multidim signals}}

In the present case we start from general formulation in Eq.~(\ref{eq: Hypothesis testing formulation})
and assume that: ($i$) $t=0$ (i.e., there is no interference, thus
$J=r$), ($ii$) $\widetilde{\bm{A}_{r}}=\bm{I}_{N}$ (thus $J=r=N)$
and ($iii$) $\widetilde{\bm{C}}\triangleq\begin{bmatrix}\bm{I}_{M} & \bm{0}_{M\times(K-M)}\end{bmatrix}$.
Such case has been dealt in \cite{Conte2003}. Therefore, the hypothesis
testing problem in canonical form\footnote{It is worth noticing that in this case the original formulation in
Eq. (\ref{eq: Hypothesis testing formulation}) is in canonical form
already.} is given by: 
\begin{equation}
\begin{cases}
\mathcal{H}_{0}: & \bm{Z}=\bm{N}\\
\mathcal{H}_{1}: & \bm{Z}=\bm{B}\,\bm{C}+\bm{N}
\end{cases}
\end{equation}
By looking at the general MIS statistic expression in Eq. (\ref{eq: MIS_final})
it is not difficult to show that the present problem admits the following
simplified partitioning:
\begin{equation}
\bm{Z}_{c}=\bm{Z}_{2},\quad\bm{Z}_{c,\perp}=\bm{Z}_{\perp,2},\quad\bm{S}_{c}=\bm{S}_{22},
\end{equation}
where $\bm{Z}_{2}\in\mathbb{C}^{N\times M}$, $\bm{Z}_{\perp,2}\in\mathbb{C}^{N\times(K-M)}$
and $\bm{S}_{22}\in\mathbb{C}^{N\times N}$, respectively. Since in
this particular setting $J=N$ holds, we exploit the alternative expression
for the MIS in Eq. (\ref{eq: MIS_final}), which shows that the MIS
reduces to a single matrix, being equal to:
\begin{align}
\bm{T}(\bm{Z}_{2},\bm{S}_{22})=\: & \bm{Z}_{2}^{\dagger}\bm{S}_{22}^{-1}\bm{Z}_{2}\nonumber \\
=\: & \bm{Z}_{2}^{\dagger}\,(\bm{Z}_{\perp,2}\bm{Z}_{\perp,2}^{\dagger})^{-1}\,\bm{Z}_{2}\label{eq: MIS_multidim signals}
\end{align}
In the latter case, it is not difficult to show that the maximal invariant
induced in the parameter space reduces to $\bm{T}_{\mathrm{p}}=\bm{B}^{\dagger}\bm{R}^{-1}\bm{B}$.

It is now interesting to compare the result in Eq. (\ref{eq: MIS_multidim signals})
with that obtained in \cite{Conte2003}. Indeed, in the aforementioned
work, the elementary action $\ell(\cdot)$ is defined as: 
\begin{gather}
\ell_{2}(\bm{Z}_{2},\bm{S}_{22})=\left(\bm{G}_{22}\,\bm{Z}_{2}\,\bm{U}_{d},\bm{G}_{22}\,\bm{S}_{22}\,\bm{G}_{22}^{\dagger}\right)\label{eq: elementary action (conte)}\\
\forall\bm{G}_{22}\in\mathcal{GL}(N)\,,\forall\bm{U}_{d}\in\mathcal{U}(M).\nonumber 
\end{gather}
which, compared to Eq. (\ref{eq: elementary action}), enforces an\emph{
additional invariance} with respect to a right subspace rotation,
via the unitary matrix $\bm{U}_{d}$. Clearly, this restricts further
the class of invariant tests. Moreover, in Eq. (\ref{eq: elementary action (conte)})
we have used $\mathcal{U}(M)$ to denote the group of unitary $M\times M$
matrices. It was shown in \cite{Conte2003} that the MIS for the elementary
action defined in Eq. (\ref{eq: elementary action (conte)}) is given
by the non-zero eigenvalues of the matrix
\begin{align}
\bm{T}_{c}\triangleq\, & \bm{S}_{22}^{-1/2}\,(\bm{Z}_{2}\bm{Z}_{2}^{\dagger})\,\bm{S}_{22}^{-1/2}\nonumber \\
=\, & (\bm{Z}_{\perp,2}\bm{Z}_{\perp,2}^{\dagger})^{-1/2}\,(\bm{Z}_{2}\bm{Z}_{2}^{\dagger})\,(\bm{Z}_{\perp,2}\bm{Z}_{\perp,2}^{\dagger})^{-1/2}\,,\label{eq: MIS_conte}
\end{align}
denoted with $\mathrm{eig}(\bm{T}_{c})$ in what follows. Remarkably,
we show hereinafter that the MIS in Eq. (\ref{eq: MIS_conte}) can
be directly linked to the expression in Eq. (\ref{eq: MIS_multidim signals}).
We first notice that, after defining $\bm{Z}_{m}\triangleq\left\{ (\bm{Z}_{\perp,2}\bm{Z}_{\perp,2}^{\dagger})^{-1/2}\,\bm{Z}_{2}\right\} \in\mathbb{C}^{N\times M}$,
the following equalities hold:
\begin{gather}
\bm{T}=(\bm{Z}_{m}^{\dagger}\,\bm{Z}_{m})\qquad\qquad\bm{T}_{c}=(\bm{Z}_{m}\,\bm{Z}_{m}^{\dagger})\label{eq: T and Tc eigs equivalence}
\end{gather}
Therefore, by construction, the matrices $\bm{T}\in\mathbb{C}^{M\times M}$
and $\bm{T}_{c}\in\mathbb{C}^{N\times N}$ are such that $\mathrm{eig}(\bm{T}_{c})=\mathrm{eig}(\bm{T})$
holds (and the vector length equals $\mathrm{min}\{M,N\}$), where
we have expressed the non-zero eigenvalues through the implicit vector-valued
function $\mathrm{eig}(\cdot)$. Then, we notice that the action $\ell_{2}(\cdot)$
can be re-interpreted as the composition of the following sub-actions:
\begin{gather}
\ell_{2,a}(\bm{Z}_{2},\bm{S}_{22})=(\bm{G}_{22}\,\bm{Z}_{2},\bm{G}_{22}\,\bm{S}_{22}\,\bm{G}_{22}^{\dagger})\quad\forall\bm{G}_{22}\in\mathcal{GL}(N)\nonumber \\
\ell_{2,b}(\bm{Z}_{2},\bm{S}_{22})=\left(\bm{Z}_{2}\bm{U}_{d},\bm{S}_{22}\right)\quad\forall\bm{U}_{d}\in\mathcal{U}(M).
\end{gather}
It is then recognized that $\ell_{2,a}(\cdot)=\ell(\cdot)$ for the
case of multidimensional signals. Previously, we have shown that the
MIS for the elementary action $\ell_{2,a}(\cdot)=\ell(\cdot)$ is
simply given by the matrix $\bm{T}$ in Eq. (\ref{eq: MIS_multidim signals}). 

Additionally, we notice that, for each $\bm{U}_{d}\in\mathcal{U}(M)$,
\begin{gather}
\bm{T}(\bar{\bm{Z}}_{2},\bar{\bm{S}}_{22})=\bm{T}(\bm{Z}_{2},\bm{S}_{22})\Rightarrow\nonumber \\
\bm{T}(\bar{\bm{Z}}_{2}\bm{U}_{d},\bar{\bm{S}}_{22})=\bm{T}(\bm{Z}_{2}\,\bm{U}_{d},\bm{S}_{22})
\end{gather}
 Now, define the action $\ell_{2,b}^{\star}(\cdot)$ as:
\begin{equation}
\ell_{2,b}^{\star}(\bm{T})=\left(\bm{U}_{d}^{\dagger}\,\bm{T}\,\bm{U}_{d}\right)\quad\forall\,\bm{U}_{d}\in\mathcal{U}(M).\label{eq: l_2,b adjoint}
\end{equation}
 where $\bm{T}\in\mathbb{H}^{M\times M}$. It is not difficult to
show that a MIS for the elementary operation $\ell_{2,b}^{\star}(\cdot)$
in Eq. (\ref{eq: l_2,b adjoint}) is given by $\mathrm{eig(}\bm{T})$.
Therefore, exploiting \cite[p. 217, Thm. 6.2.2]{Lehmann2006}, it
follows that a MIS for the action $\ell_{2}(\cdot)$ is the composite
function $\mathrm{eig}(\bm{T}(\bm{Z}_{2},\bm{S}_{22}))$. However,
since as underlined in Eq. (\ref{eq: T and Tc eigs equivalence}),
we have $\mathrm{eig}(\bm{T})=\mathrm{eig}(\bm{T}_{c})$, this clearly
coincides with the result in \cite{Conte2003}.

Finally, by similar reasoning it is not difficult to show that, in
such a case, the induced maximal invariant is given by $\mathrm{eig}(\bm{T}_{\mathrm{p}})=\mathrm{eig}(\bm{B}^{\dagger}\bm{R}^{-1}\bm{B})=\mathrm{eig}(\bm{R}^{-1/2}\bm{B}\,\bm{B}^{\dagger}\bm{R}^{-1/2})$,
thus obtaining the result in \cite{Conte2003}.

\subsection{Range-spread Targets \label{sub: MIS - Range spread Targets}}

In the present case we start from general formulation in Eq.~(\ref{eq: Hypothesis testing formulation})
and assume that: ($i$) $t=0$ (i.e., there is no interference, thus
$J=r$); ($ii$) $r=1$, thus the matrices $\widetilde{\bm{A}}_{r}$
and $\widetilde{\bm{B}}_{r}$ collapse to $\widetilde{\bm{a}_{r}}\in\mathbb{C}^{N\times1}$
and $\widetilde{\bm{b}}_{r}\in\mathbb{C}^{1\times M}$ (i.e. a row
vector), respectively; ($iii$) $\widetilde{\bm{C}}\triangleq\begin{bmatrix}\bm{I}_{M} & \bm{0}_{M\times K-M}\end{bmatrix}$.
Such case has been dealt in \cite{Conte2001,Raghavan2013}. Therefore,
the hypothesis testing in canonical form is given by: 
\begin{equation}
\begin{cases}
\mathcal{H}_{0}: & \bm{Z}=\bm{N}\\
\mathcal{H}_{1}: & \bm{Z}=\bm{a}\,\bm{b}\,\bm{C}+\bm{N}
\end{cases}
\end{equation}
where $\bm{a}\triangleq\begin{bmatrix}1 & 0 & \cdots & 0\end{bmatrix}^{T}\in\mathbb{C}^{N\times1}$,
$\bm{b}\in\mathbb{C}^{1\times M}$ and $\bm{C=}\widetilde{\bm{C}}$,
respectively. By looking at the general MIS expression in Eq. (\ref{eq: MIS_final}),
it is not difficult to show that the present problem admits the following
simplified partitioning:
\begin{equation}
\bm{Z}_{c}=\begin{bmatrix}\bm{z}_{2}\\
\bm{Z}_{3}
\end{bmatrix},\quad\bm{Z}_{c,\perp}=\begin{bmatrix}\bm{z}_{\perp,2}\\
\bm{Z}_{\perp,3}
\end{bmatrix},\quad\bm{S}_{c}=\begin{bmatrix}s_{22} & \bm{s}_{23}\\
\bm{s}_{32} & \bm{S}_{33}
\end{bmatrix},
\end{equation}
where $\bm{z}_{2}\in\mathbb{C}^{1\times M}$ (i.e., a row vector),
$\bm{Z}_{3}\in\mathbb{C}^{(N-1)\times M}$, $\bm{Z}_{\perp,2}\in\mathbb{C}^{1\times(K-M)}$
(i.e., a row vector), $\bm{Z}_{\perp,3}\in\mathbb{C}^{(N-1)\times(K-M)}$,
$s_{22}\in\mathbb{C}$, $\bm{s}_{23}\in\mathbb{C}^{1\times(N-1)}$
(i.e., a row vector), $\bm{s}_{32}\in\mathbb{C}^{(N-1)\times1}$ and
$\bm{S}_{33}\in\mathbb{C}^{(N-1)\times(N-1)}$, respectively. Exploiting
the above partitioning, gives $s_{2.3}=(s_{22}-\bm{s}_{23}\,\bm{S}_{33}^{-1}\,\bm{s}_{32})\in\mathbb{C}$
(i.e., a scalar) and the simplified expression:
\begin{equation}
\bm{z}_{2.3}=(\bm{z}_{2}-\bm{s}_{23}\,\bm{S}_{33}^{-1}\,\bm{Z}_{3})\in\mathbb{C}^{1\times M}
\end{equation}
Given the simplified expressions for $\bm{z}_{2.3}$ (row vector)
and $s_{2.3}$ (scalar), it is not difficult to show that the two
matrix components of the MIS are given by:
\begin{gather}
\bm{T}_{a}=\left(\frac{1}{s_{2.3}}\right)\bm{z}_{2.3}^{\dagger}\bm{z}_{2.3}\qquad\bm{T}_{b}=\bm{Z}_{3}^{\dagger}\,\bm{S}_{33}^{-1}\,\bm{Z}_{3}\label{eq: MIS range_spread}
\end{gather}
where the matrix $\bm{T}_{a}$ is rank-one in this specific case (as
it is the output of a dyadic product). Also, the induced maximal invariant
in the parameter space equals $\bm{T}_{\mathrm{p}}=(\frac{1}{r_{2.3}})\,\bm{b}^{\dagger}\bm{b}$,
i.e., a rank-one matrix.

It is now of interest comparing the MIS represented by Eq.~(\ref{eq: MIS range_spread})
with that obtained in \cite{Raghavan2013}. The approach taken in
the following is similar to that used for multidimensional signals
in Sec. \ref{sub: MIS multidim signals}. However, due to the more
tedious mathematics involved, we confine the proof to Appendix \ref{sec: Appendix _  Range-spread Raghavan}
and we only state the results hereinafter.

Indeed, in the aforementioned work, the elementary action $\ell(\cdot)$
is defined as: 
\begin{gather}
\ell_{2}(\bm{Z}_{c},\bm{S}_{c})=\left(\bm{G}\,\bm{Z}_{c}\,\bm{U}_{d},\bm{G}\,\bm{S}_{c}\,\bm{G}^{\dagger}\right)\,,\label{eq: elementary action (raghavan)}\\
\forall\,\bm{G}\in\mathcal{G}\,,\quad\forall\,\bm{U}_{d}\in\mathcal{U}(M),\nonumber 
\end{gather}
which, compared to Eq. (\ref{eq: elementary action}), enforces an
additional invariance with respect to a right subspace rotation of
primary data, via the unitary matrix $\bm{U}_{d}$. It was shown in
\cite{Raghavan2013} that the MIS for the elementary action defined
in Eq. (\ref{eq: elementary action (raghavan)}) is given by the eigenvalues
of the matrices
\begin{gather}
(\bm{T}_{a}+\bm{T}_{b}),\quad\quad\bm{T}_{b},\label{eq: MIS_raghavan2013}
\end{gather}
denoted with $\mathrm{eig}(\bm{T}_{a}+\bm{T}_{b})$ and $\mathrm{eig}(\bm{T}_{b})$
in what follows. Remarkably, we show hereinafter that the MIS in Eq.
(\ref{eq: MIS_raghavan2013}) can be directly linked to the expression
in Eq. (\ref{eq: MIS range_spread}). We first notice that the action
$\ell_{2}(\cdot)$ can be re-interpreted as the composition of the
following sub-actions:
\begin{gather}
\ell_{2,a}(\bm{Z}_{c},\bm{S}_{c})=(\bm{G}\,\bm{Z}_{c},\bm{G}\,\bm{S}_{c}\,\bm{G}^{\dagger}),\quad\forall\bm{G}\in\mathcal{G},\nonumber \\
\ell_{2,b}(\bm{Z}_{c},\bm{S}_{c})=\left(\bm{Z}_{c}\,\bm{U}_{d},\bm{S}_{c}\right),\quad\forall\,\bm{U}_{d}\in\mathcal{U}(M).
\end{gather}
It is then recognized that $\ell_{2,a}(\cdot)=\ell(\cdot)$ for the
case of range-spread signals. Also, we have previously shown that
a MIS for the elementary action $\ell_{2,a}(\cdot)=\ell(\cdot)$ is
given by Eq. (\ref{eq: MIS range_spread}).

Additionally, we notice that, for each $\bm{U}_{d}\in\mathcal{U}(M)$,
\begin{gather}
\bm{T}(\bar{\bm{Z}}_{c},\bar{\bm{S}}_{c})=\bm{T}(\bm{Z}_{c},\bm{S}_{c})\Rightarrow\nonumber \\
\bm{T}(\bar{\bm{Z}}_{c}\,\bm{U}_{d},\bar{\bm{S}}_{c})=\bm{T}(\bm{Z}_{c}\,\bm{U}_{d},\bm{S}_{c})
\end{gather}
 Now, define the action $\ell_{2,b}^{\star}(\cdot)$ as:
\begin{equation}
\ell_{2,b}^{\star}(\bm{T}_{a},\bm{T}_{b})=\left(\bm{U}_{d}^{\dagger}\,\bm{T}_{a}\,\bm{U}_{d},\bm{U}_{d}^{\dagger}\,\bm{T}_{b}\,\bm{U}_{d}\right),\label{eq: l_2,b adjoint - range spread}
\end{equation}
$\forall\,\bm{U}_{d}\in\mathcal{U}(M)$, where $\bm{T}_{b}\in\mathbb{H}^{M\times M}$
and $\bm{T}_{a}=(\bm{a}\bm{a}^{\dagger})$ (that is, a rank-one matrix).
It is shown in Appendix \ref{sec: Appendix _  Range-spread Raghavan}
that the MIS for the elementary operation $\ell_{2,b}^{\star}(\cdot)$
in Eq.~(\ref{eq: l_2,b adjoint - range spread}) is given by $\{\mathrm{eig}(\bm{T}_{b}),\mathrm{eig(}\bm{T}_{a}+\bm{T}_{b})\}$.
Therefore, by exploiting \cite[p. 217, Thm. 6.2.2]{Lehmann2006},
it follows that the MIS for the action $\ell_{2}(\cdot)$ is the composite
function 
\begin{gather}
\begin{cases}
\mathrm{eig}\left(\bm{Z}_{3}^{\dagger}\,\bm{S}_{33}^{-1}\,\bm{Z}_{3}\right)\\
\mathrm{eig}\left(\bm{Z}_{3}^{\dagger}\,\bm{S}_{33}^{-1}\,\bm{Z}_{3}+\left(\frac{1}{s_{2.3}}\right)\bm{z}_{2.3}^{\dagger}\bm{z}_{2.3}\right)
\end{cases},
\end{gather}
which clearly coincides with the result in \cite{Raghavan2013}. 

Finally, it is not difficult to show that the induced maximal invariant
in such a case can be obtained as $\mathrm{eig(}\bm{T}_{\mathrm{p}})=(\frac{\left\Vert \bm{b}\right\Vert ^{2}}{r_{2.3}})=\left\Vert \bm{b}\right\Vert ^{2}(\bm{a}^{\dagger}\bm{R}^{-1}\bm{a})$
(since the rank-one induced maximal invariant has only one non-zero
eigenvalue), which represents the overall SINR over the $M$ cells,
as defined in \cite{Raghavan2013}.

\subsection{Standard GMANOVA}

Finally, in the present case we start from general formulation in
Eq.~(\ref{eq: Hypothesis testing formulation}) and assume that:
($i$) $t=0$ (i.e., there is no interference, thus $J=r$). Such
model clearly coincides with that analyzed in \cite{Kelly1989,Liu2014},
\emph{unfortunately not dealing with the derivation of the MIS}. Therefore,
the hypothesis testing in canonical form is given by: 
\begin{equation}
\begin{cases}
\mathcal{H}_{0}: & \bm{Z}=\bm{N}\\
\mathcal{H}_{1}: & \bm{Z}=\bm{A}\,\bm{B}\,\bm{C}+\bm{N}
\end{cases}
\end{equation}
where $\bm{A}=\begin{bmatrix}\bm{I}_{J} & \bm{0}_{J\times(N-J)}\end{bmatrix}^{T}$
and $\bm{B}\in\mathbb{C}^{J\times M}$, respectively. By looking at
the general MIS statistic expression in Eq. (\ref{eq: MIS_final})
it is not difficult to show that the present problem admits the following
simplified partitioning:
\begin{equation}
\bm{Z}_{c}=\begin{bmatrix}\bm{Z}_{2}\\
\bm{Z}_{3}
\end{bmatrix},\;\bm{Z}_{c,\perp}=\begin{bmatrix}\bm{Z}_{\perp,2}\\
\bm{Z}_{\perp,3}
\end{bmatrix},\;\bm{S}_{c}=\begin{bmatrix}\bm{S}_{22} & \bm{S}_{23}\\
\bm{S}_{32} & \bm{S}_{33}
\end{bmatrix},\label{eq: Simplified partitioning MIS (GMANOVA no interf)}
\end{equation}
where $\bm{Z}_{2}\in\mathbb{C}^{J\times M}$, $\bm{Z}_{3}\in\mathbb{C}^{(N-J)\times M}$,
$\bm{Z}_{\perp,2}\in\mathbb{C}^{J\times(K-M)}$, $\bm{Z}_{\perp,3}\in\mathbb{C}^{(N-J)\times(K-M)}$,
$\bm{S}_{22}\in\mathbb{C}^{J\times J}$, $\bm{S}_{23}\in\mathbb{C}^{J\times(N-J)}$,
$\bm{S}_{32}\in\mathbb{C}^{(N-J)\times J}$ and $\bm{S}_{33}\in\mathbb{C}^{(N-J)\times(N-J)}$,
respectively. Given the simplified definitions in Eq. (\ref{eq: Simplified partitioning MIS (GMANOVA no interf)}),
the MIS is readily obtained via the standard formula in Eq. (\ref{eq: MIS_final}). 

Finally, the induced maximal invariant is obtained through the standard
formula $\bm{T}_{\mathrm{p}}=(\bm{B}^{\dagger}\,\bm{R}_{2.3}^{-1}\,\bm{B})$.
The sole difference consists in the rank of matrix $\bm{T}_{\mathrm{p}}$,
being equal to $\min\{J,M\}$, i.e., there is no reduction in the
observation space due to structured interference.

\section{Conclusions\label{sec: Conclusions}}

In the first part of this work, we have studied a generalization of
GMANOVA model (denoted as I-GMANOVA) which comprises additional (deterministic)
structured interference, modeling possible jamming interference. The
study has been conducted with the help of the statistical theory of
invariance. For the present problem, the group of trasformations leaving
the hypothesis testing problem invariant was derived, thus allowing
identification of trasformations which enforce CFARness. Then, a MIS
was derived for the aforementioned group, thus explicitly underlining
the basic structure of a generic CFAR receiver (several examples of
CFAR receivers, based on theoretically-founded criteria, will be derived
in part II of the present work). 

Furthermore, a statistical characterization of the considered MIS
under both hypotheses was obtained, thus allowing for an efficient
stochastic representation. As a byproduct, the general form of the
induced maximal invariant in the parameter space was obtained for
the considered hypothesis testing. Finally, the general MIS expression
was particularized and compared with MIS obtained in specific instances
found in the open literature. Analogies to other expressions of the
MIS, obtained by enforcing invariance to a wider class of transformations
(cf. Sec. \ref{sub: MIS - Range spread Targets} and \ref{sub: MIS multidim signals}),
were underlined and discussed.

\appendices{}

\section{Invariance of the problem\protect \\
 with respect to the group $\mathcal{L}$\label{sec: Invariance wrt group L}}

In this appendix we prove the invariance of the hypothesis testing
problem in (\ref{eq: Transformed data - hypothesis testing problem})
with respect to the group of trasformations $\mathcal{L}$ defined
in Sec. \ref{sub: Desired Invariance Properties}. Let ($\bm{G},\bm{F})\in\mathcal{L}$
and observe that, under $\mathcal{H}_{1}$, the columns of $\bm{G}\bm{Z}_{c}+\bm{F}$
are independent complex normal vectors with covariance matrix $\bm{G}\bm{R}\bm{G}^{\dagger}$
and mean:
\begin{align}
\bm{G}\bm{A}\bm{B}_{s}+\bm{F} & =\begin{bmatrix}\bm{G}_{11}\bm{B}_{t,1}+\bm{G}_{12}\bm{B}+\bm{F}_{1}\\
\bm{G}_{22}\bm{B}\\
\bm{0}_{(N-J)\times M}
\end{bmatrix}\\
 & =\begin{bmatrix}\bm{B}_{t,1}^{'}\\
\bm{B}^{'}\\
\bm{0}_{(N-J)\times M}
\end{bmatrix}=\bm{A}\bm{B}_{s}^{'}
\end{align}
 where we have employed the definitions $\bm{B}_{t,1}^{'}\triangleq(\bm{G}_{11}\bm{B}_{t,1}+\bm{G}_{12}\bm{B}+\bm{F}_{11})\in\mathbb{C}^{t\times M}$
and $\bm{B}^{'}\triangleq(\bm{G}_{22}\bm{B})\in\mathbb{C}^{r\times M}$,
respectively. Also, aiming at compact notation, we have denoted $\bm{B}_{s}^{'}\triangleq\begin{bmatrix}(\bm{B}_{t,1}^{'})^{T} & (\bm{B}^{'})^{T}\end{bmatrix}^{T}$.
Furthermore, it is not difficult to show that $\bm{G}\bm{S}_{c}\bm{G}^{\dagger}=(\bm{G}\bm{Z}_{c,\perp})(\bm{G}\bm{Z}_{c,\perp})^{\dagger}$,
with $(\bm{G}\bm{Z}_{c,\perp})\sim\mathcal{N}_{\mathbb{C}}(\bm{0}_{N\times(K-M)},\,\bm{I}_{K-M},\bm{G}\,\bm{R}\,\bm{G}^{\dagger})$.

On the other hand, when $\mathcal{H}_{0}$ holds true, $\bm{G}\bm{Z}_{c}+\bm{F}$
shares the same covariance structure as in the case of $\mathcal{H}_{1},$
except for the mean, which becomes
\begin{align}
\bm{G}\bm{A}\,\begin{bmatrix}\bm{B}_{t,0}\\
\bm{0}_{r\times M}
\end{bmatrix}+\bm{F} & =\begin{bmatrix}\bm{G}_{11}\bm{B}_{t,0}+\bm{F}_{1}\\
\bm{0}_{r\times M}\\
\bm{0}_{(N-J)\times M}
\end{bmatrix}\\
 & =\bm{A}\,\begin{bmatrix}\bm{B}_{t,0}^{'}\\
\bm{0}_{r\times M}
\end{bmatrix}
\end{align}
where $\bm{B}_{t,0}^{'}\triangleq(\bm{G}_{11}\bm{B}_{t,0}+\bm{F}_{1})\in\mathbb{C}^{t\times M}$.
Again, it is not difficult to show that $(\bm{G}\bm{Z}_{c,\perp})\sim\mathcal{CN}_{N\times(K-M)}(\bm{0}_{N\times(K-M)},\,\bm{I}_{K-M},\bm{G}\,\bm{R}\,\bm{G}^{\dagger})$.

Therefore, it is apparent that the original partition of the parameter
space, the data distribution, and the structure of the subspace containing
the useful signal components are preserved after the transformation
$(\bm{G},\bm{F})$. Indeed, the following equivalence holds between
the original and the trasformed test: 
\begin{equation}
\begin{cases}
\mathcal{H}_{0}\,:\, & ||\bm{B}||_{F}=0\,\Longleftrightarrow||\bm{B}^{'}||_{F}=0,\\
\mathcal{H}_{1}\,:\, & ||\bm{B}||_{F}>0\,\Longleftrightarrow||\bm{B}^{'}||_{F}>0,
\end{cases}
\end{equation}
where the nuisance parameters in the transformed space are $\bm{B}_{t,i}^{'}$
and $(\bm{G}\,\bm{R}\,\bm{G}^{\dagger})$.

\section{Derivation of the Maximal Invariant Statistic \label{sec: Appendix _ MIS derivation}}

In the present appendix we provide a proof for Prop.~\ref{prop: Maximal Invariant Statistic}.
In particular, hereinafter we will focus on the case $J<N$, as to
the derivation for $J=N$ can be obtained through identical steps.
Before proceeding further, we recall that a statistic $\bm{T}(\bm{Z}_{c},\bm{S}_{c})$
is said to be a maximal invariant with respect to the group of transformations
$\mathcal{L}$ iff 
\begin{align}
(\mathrm{a})\quad\bm{T}(\bm{Z}_{c},\bm{S}_{c}) & =\bm{T}[\ell(\bm{Z}_{c},\bm{S}_{c})],\quad\forall\ell\in\mathcal{L}\,;\\
(\mathrm{b})\quad\bm{T}(\bm{Z}_{c},\bm{S}_{c}) & =\bm{T}(\bar{\bm{Z}}_{c},\bar{\bm{S}}_{c})\Rightarrow\nonumber \\
\exists\,\ell & \in\mathcal{L}\,:\,(\bm{Z}_{c},\bm{S}_{c})=\ell(\bar{\bm{Z}}_{c},\bar{\bm{S}}_{c})\,.
\end{align}
Conditions ($\mathrm{a}$) and ($\mathrm{b}$) correspond to the so-called
\emph{invariance} and \emph{maximality} properties, respectively.
In order to prove $(\mathrm{a})$, we first consider the following
partitioning of matrix $\bm{G}$ and sub-matrix of $\bm{S}_{c}$:
\begin{equation}
\bm{G}=\begin{bmatrix}\bm{G}_{1} & \bm{G}_{2}\\
\bm{0}_{(N-t)\times t} & \bm{G}_{3}
\end{bmatrix}\,,\quad\bm{S}_{2}\triangleq\begin{bmatrix}\bm{S}_{22} & \bm{S}_{23}\\
\bm{S}_{32} & \bm{S}_{33}
\end{bmatrix},\label{eq: G-S_2 definition}
\end{equation}
where $\bm{S}_{2}\in\mathbb{C}^{(N-t)\times(N-t)}$, $\bm{G}_{1}\triangleq\bm{G}_{11}\in\mathbb{C}^{t\times t}$,
$\bm{G}_{2}\triangleq\begin{bmatrix}\bm{G}_{12} & \bm{G}_{13}\end{bmatrix}\in\mathbb{C}^{t\times(N-t)}$
and 
\begin{equation}
\bm{G}_{3}\triangleq\begin{bmatrix}\bm{G}_{22} & \bm{G}_{23}\\
\bm{0}_{(N-J)\times r} & \bm{G}_{33}
\end{bmatrix}\in\mathbb{C}^{(N-t)\times(N-t)}.\label{eq: Appendix_block triangular G3}
\end{equation}
 Then, let $(\bar{\bm{Z}}_{c},\bar{\bm{S}}_{c})\triangleq\ell(\bm{Z}_{c},\bm{S}_{c})$,
with
\begin{equation}
\bar{\bm{Z}}_{c}=\bm{G}\bm{Z}_{c}+\bm{F},\quad\quad\bar{\bm{S}}_{c}=\bm{G}\,\bm{S}_{c}\,\bm{G}^{\dagger}.
\end{equation}
It is apparent that the following equalities hold, when exploiting
the specific structure of $\bm{G}$ and $\bm{F}$:
\begin{eqnarray}
\bar{\bm{Z}}_{2} & = & \bm{G}_{22}\,\bm{Z}_{2}+\bm{G}_{23}\,\bm{Z}_{3},\\
\bar{\bm{Z}}_{3} & = & \bm{G}_{33}\,\bm{Z}_{3},\label{eq: Z3 bar}
\end{eqnarray}
and 
\begin{equation}
\bar{\bm{S}}_{2}=\bm{G}_{3}\,\bm{S}_{2}\,\bm{G}_{3}^{\dagger}\,,\label{eq: S2_bar}
\end{equation}
where $\bar{\bm{S}}_{2}$ is similarly defined as in Eq. (\ref{eq: G-S_2 definition}).
From Eq. (\ref{eq: S2_bar}), it can be inferred that:
\begin{eqnarray}
\bar{\bm{S}}_{22} & = & \begin{bmatrix}\bm{G}_{22} & \bm{G}_{23}\end{bmatrix}\,\bm{S}_{2}\,\begin{bmatrix}\bm{G}_{22} & \bm{G}_{23}\end{bmatrix}^{\dagger}\\
\bar{\bm{S}}_{23} & = & \bm{G}_{22}\,\bm{S}_{23}\,\bm{G}_{33}^{\dagger}+\bm{G}_{23}\,\bm{S}_{33}\,\bm{G}_{33}^{\dagger}\\
\bar{\bm{S}}_{33} & = & \bm{G}_{33}\,\bm{S}_{33}\,\bm{G}_{33}^{\dagger}\label{eq: S3 bar}
\end{eqnarray}
Additionally, exploiting the appropriate substitutions, it can be
shown that:
\begin{gather}
\bar{\bm{Z}}_{2.3}=\left(\bar{\bm{Z}}_{2}-\bar{\bm{S}}_{23}\,\bar{\bm{S}}_{33}^{-1}\,\bar{\bm{Z}}_{3}\right)=\bm{G}_{22}\,\bm{Z}_{2.3}\label{eq: Z2.3 bar}\\
\bar{\bm{S}}_{2.3}=\left(\bar{\bm{S}}_{22}-\bar{\bm{S}}_{23}\,\bar{\bm{S}}_{33}^{-1}\,\bar{\bm{S}}_{32}\right)=\bm{G}_{22}\,\bm{S}_{2.3}\,\bm{G}_{22}^{\dagger}\label{eq: S 2.3 bar}
\end{gather}
Finally, substituting Eqs. (\ref{eq: Z3 bar}), (\ref{eq: S3 bar}),
(\ref{eq: Z2.3 bar}), and (\ref{eq: S 2.3 bar}) into (\ref{eq: MIS_final}),
we obtain:
\begin{gather}
\bm{T}(\ell(\bm{Z}_{c},\bm{S}_{c}))=\begin{bmatrix}\bar{\bm{Z}}_{2.3}^{\dagger}\,\bar{\bm{S}}_{2.3}^{-1}\,\bar{\bm{Z}}_{2.3}\\
\bar{\bm{Z}}_{3}^{\dagger}\,\bar{\bm{S}}_{33}^{-1}\,\bar{\bm{Z}}_{3}
\end{bmatrix}\\
=\begin{bmatrix}\bm{Z}_{2.3}^{\dagger}\,\bm{G}_{22}^{\dagger}\,(\bm{G}_{22}^{\dagger})^{-1}\,\bm{S}_{2.3}^{-1}\,\bm{G}_{22}^{-1}\,\bm{G}_{22}\,\bm{Z}_{2.3}\\
\bm{Z}_{3}^{\dagger}\,\bm{G}_{33}^{\dagger}\,(\bm{G}_{33}^{\dagger})^{-1}\,\bm{S}_{33}^{-1}\,(\bm{G}_{33}^{-1})\,\bm{G}_{33}\,\bm{Z}_{3}
\end{bmatrix}\\
=\begin{bmatrix}\bm{Z}_{2.3}^{\dagger}\,\bm{S}_{2.3}^{-1}\,\bm{Z}_{2.3}\\
\bm{Z}_{3}^{\dagger}\,\bm{S}_{33}^{-1}\,\bm{Z}_{3}
\end{bmatrix}
\end{gather}
which thus proves $(\mathrm{a})$. 

Now, in order to prove $(\mathrm{b})$, assume that:
\begin{eqnarray}
\bm{T}(\bm{Z}_{c},\bm{S}_{c}) & = & \bm{T}(\bar{\bm{Z}}_{c},\bar{\bm{S}}_{c})\\
\begin{bmatrix}\bm{Z}_{2.3}^{\dagger}\,\bm{S}_{2.3}^{-1}\,\bm{Z}_{2.3}\\
\bm{Z}_{3}^{\dagger}\,\bm{S}_{33}^{-1}\,\bm{Z}_{3}
\end{bmatrix} & = & \begin{bmatrix}\bar{\bm{Z}}_{2.3}^{\dagger}\,\bar{\bm{S}}_{2.3}^{-1}\,\bar{\bm{Z}}_{2.3}\\
\bar{\bm{Z}}_{3}^{\dagger}\,\bar{\bm{S}}_{33}^{-1}\,\bar{\bm{Z}}_{3}
\end{bmatrix}
\end{eqnarray}
The last equality can be recast as the following pair of equalities
\begin{gather}
\bm{Y}_{2.3}\,\bm{Y}_{2.3}^{\dagger}=\bar{\bm{Y}}_{2.3}\,\bar{\bm{Y}}_{2.3}^{\dagger},\quad\quad\bm{Y}_{3}\,\bm{Y}_{3}^{\dagger}=\bar{\bm{Y}}_{3}\,\bar{\bm{Y}}_{3}^{\dagger},\label{eq: proof (b) MIS - pair no 1}
\end{gather}
where $\bm{Y}_{2.3}\triangleq(\bm{S}_{2.3}^{-1/2}\bm{Z}_{2.3})^{\dagger}$,
$\bar{\bm{Y}}_{2.3}\triangleq(\bar{\bm{S}}_{2.3}^{-1/2}\bar{\bm{Z}}_{2.3})^{\dagger}$,
$\bm{Y}_{3}\triangleq(\bm{S}_{33}^{-1/2}\bm{Z}_{3})^{\dagger}$ and
$\bar{\bm{Y}}_{3}=(\bar{\bm{S}}_{33}^{-1/2}\bar{\bm{Z}}_{3})^{\dagger}$.
It follows from direct inspection of Eq. (\ref{eq: proof (b) MIS - pair no 1})
that there exist unitary matrices\footnote{Such property can be verified as follows: given the equality $\bm{A}\bm{A}^{\dagger}=\bm{B}\bm{B}^{\dagger}$
between two generic matrices $\bm{A}$ and $\bm{B}$ and, after defining
the eigenvalue decompositions $(\bm{A}\bm{A}^{\dagger})=\bm{U}_{A}\,\bm{\Lambda}_{A}\,\bm{U}_{A}^{\dagger}$
and $(\bm{B}\bm{B}^{\dagger})=\bm{U}_{B}\,\bm{\Lambda}_{B}\,\bm{U}_{B}^{\dagger}$,
and the SVDs $\bm{A}=\bm{U}_{A}\,\bm{\Sigma}_{A}\,\bm{V}_{A}^{\dagger}$
and $\bm{B}=\bm{U}_{B}\,\bm{\Sigma}_{B}\,\bm{V}_{B}^{\dagger}$, it
is apparent that such equality implies: ($i$) $\bm{\Sigma}_{A}=\bm{\Sigma}_{B}\widetilde{\bm{D}}$;
($ii$) $\bm{U}_{A}=\bm{U}_{B}\,\bm{D}$, where $\bm{D}$ and $\widetilde{\bm{D}}$
denote diagonal matrices of phasors (recall that $\bm{\Lambda}_{A}=\bm{\Sigma}_{A}\bm{\Sigma}_{A}^{*}$and
$\bm{\Lambda}_{B}=\bm{\Sigma}_{B}\bm{\Sigma}_{B}^{*}$). Thus, it
follows from substitution that $\bm{A}=\bm{U}_{B}\bm{D}\,\bm{\Sigma}_{B}\,\widetilde{\bm{D}}\,\bm{V}_{A}^{\dagger}=\bm{U}_{B}\,\bm{\Sigma}_{B}\,\widetilde{\bm{D}}\,\bm{D}\,\bm{V}_{A}^{\dagger}$.
Therefore, after defining the unitary matrix $\bm{U}^{\star}\triangleq\bm{V}_{A}\,(\widetilde{\bm{D}}\,\bm{D})^{*}\bm{U}_{B}^{\dagger}$,
it is finally demonstrated that $\bm{A}\bm{U}^{\star}=\bm{B}$.} $\bm{U}_{2.3}\in\mathbb{C}^{r\times r}$ and $\bm{U}_{3}\in\mathbb{C}^{(N-J)\times(N-J)}$
such that $\bm{Y}_{2.3}=\bar{\bm{Y}}_{2.3}\,\bm{U}_{2.3}$ and $\bm{Y}_{3}=\bar{\bm{Y}}_{3}\,\bm{U}_{3}$.

First, let us define the following block-triangular decompositions
for matrices $\bm{S}_{2}=\bm{L}_{2}^{\dagger}\,\bm{L}_{2}$ and $\bar{\bm{S}}_{2}=\bar{\bm{L}}_{2}^{\dagger}\,\bar{\bm{L}}_{2}$,
where:
\begin{align}
\bm{L}_{2} & \triangleq\begin{bmatrix}\bm{S}_{2.3}^{1/2} & \bm{0}_{r\times(N-J)}\\
\bm{S}_{33}^{-1/2}\bm{S}_{32} & \bm{S}_{33}^{1/2}
\end{bmatrix}\\
\bar{\bm{L}}_{2} & \triangleq\begin{bmatrix}\bar{\bm{S}}_{2.3}^{1/2} & \bm{0}_{r\times(N-J)}\\
\bar{\bm{S}}_{33}^{-1/2}\bar{\bm{S}}_{32} & \bar{\bm{S}}_{33}^{1/2}
\end{bmatrix}
\end{align}
Therefore, given the aforementioned definitions, it can be shown that:
\begin{align}
\begin{bmatrix}\bm{Y}_{2.3}^{\dagger}\\
\bm{Y}_{3}^{\dagger}
\end{bmatrix} & =(\bm{L}_{2}^{\dagger})^{-1}\bm{Z}_{23}=\label{eq: Y23 resolving (a)}\\
\begin{bmatrix}\bm{U}_{2.3}^{\dagger}\,\bar{\bm{Y}}_{2.3}^{\dagger}\\
\bm{U}_{3}^{\dagger}\,\bar{\bm{Y}}_{3}^{\dagger}
\end{bmatrix} & =\bm{U}_{1}(\bar{\bm{L}}_{2}^{\dagger})^{-1}\bar{\bm{Z}}_{23}\label{eq: Y23 resolving (b)}
\end{align}
where $\bm{Z}_{23}\triangleq\begin{bmatrix}\bm{Z}_{2}^{T} & \bm{Z}_{3}^{T}\end{bmatrix}^{T}$
and $\bm{U}_{1}\triangleq\mathrm{diag}(\bm{U}_{2.3}^{\dagger},\bm{U}_{3}^{\dagger})$,
respectively. From comparison of right hand side. of Eqs. (\ref{eq: Y23 resolving (a)})
and (\ref{eq: Y23 resolving (b)}), it readily follows that 
\begin{equation}
\bm{Z}_{23}=\bm{L}_{2}^{\dagger}\,\bm{U}_{1}(\bar{\bm{L}}_{2}^{\dagger})^{-1}\bar{\bm{Z}}_{23}\,.\label{eq: Z23 MIS}
\end{equation}
From inspection of Eq. (\ref{eq: Z23 MIS}), it is apparent that selecting
the transformation $\bm{G}_{3}=\bm{L}_{2}^{\dagger}\,\bm{U}_{1}(\bar{\bm{L}}_{2}^{\dagger})^{-1}$
(which is block-triangular as dictated by Eq. (\ref{eq: Appendix_block triangular G3}))
automatically verifies the set of equations:
\begin{gather}
\begin{cases}
(i) & \bm{G}_{3}\,\bar{\bm{Z}}_{23}=\bm{Z}_{23}\\
(ii) & \bm{G}_{3}\,\bar{\bm{S}}_{2}\,\bm{G}_{3}^{\dagger}=\bm{S}_{2}
\end{cases}
\end{gather}
since it also holds 
\begin{gather}
(\bm{L}_{2}^{\dagger})\bm{U}_{1}(\bar{\bm{L}}_{2}^{\dagger})^{-1}\bar{\bm{S}}_{2}(\bar{\bm{L}}_{2})^{-1}\bm{U}_{1}^{\dagger}\bm{L}_{2}=\nonumber \\
(\bm{L}_{2}^{\dagger})\bm{U}_{1}(\bar{\bm{L}}_{2}^{\dagger})^{-1}\,\bar{\bm{L}}_{2}^{\dagger}\,\bar{\bm{L}}_{2}\,(\bar{\bm{L}}_{2})^{-1}\bm{U}_{1}^{\dagger}\bm{L}_{2}=\bm{L}_{2}^{\dagger}\,\bm{L}_{2}=\bm{S}_{2}
\end{gather}
Finally, we shown how to build the remaining blocks of $\bm{G}$.
To this end, let us denote $\bm{S}_{3}\triangleq\begin{bmatrix}\bm{S}_{12} & \bm{S}_{13}\end{bmatrix}$,
$\bar{\bm{S}}_{3}\triangleq\begin{bmatrix}\bar{\bm{S}}_{12} & \bar{\bm{S}}_{13}\end{bmatrix}$,
$\bm{S}_{1}\triangleq\bm{S}_{11}$, and $\bar{\bm{S}}_{1}\triangleq\bar{\bm{S}}_{11}$,
and consider the block-triangular decompositions for matrices $\bm{S}_{c}=\bm{L}_{c}^{\dagger}\,\bm{L}_{c}$
and $\bar{\bm{S}}_{c}=\bar{\bm{L}}_{c}^{\dagger}\,\bar{\bm{L}}_{c}$
as:
\begin{align}
\bm{L}_{c} & \triangleq\begin{bmatrix}\bm{S}_{.1}^{1/2} & \bm{0}_{t\times(N-t)}\\
(\bm{L}_{2}^{\dagger}\bm{U}_{1})^{-1}\bm{S}_{3}^{\dagger} & (\bm{U}_{1}^{\dagger}\bm{L}_{2})
\end{bmatrix}\\
\bar{\bm{L}}_{c} & \triangleq\begin{bmatrix}\bar{\bm{S}}_{.1}^{1/2} & \bm{0}_{t\times(N-t)}\\
(\bar{\bm{L}}_{2}^{\dagger})^{-1}\bar{\bm{S}}_{3}^{\dagger} & \bar{\bm{L}}_{2}
\end{bmatrix}
\end{align}
where $\bm{S}_{.1}\triangleq\bm{S}_{1}-\bm{S}_{3}\bm{S}_{2}^{-1}\bm{S}_{3}^{\dagger}\in\mathbb{C}^{t\times t}$
and $\bar{\bm{S}}_{.1}\triangleq\bar{\bm{S}}_{1}-\bar{\bm{S}}_{3}\bar{\bm{S}}_{2}^{-1}\bar{\bm{S}}_{3}^{\dagger}\in\mathbb{C}^{t\times t}$.
Also, since we need to ensure $(\bm{G}\,\bar{\bm{S}}_{c}\,\bm{G}^{\dagger})=\bm{S}_{c}$,
it suffices that 
\begin{gather}
\bar{\bm{L}}_{c}\,\bm{G}^{\dagger}=\bm{L}_{c}\\
\begin{bmatrix}\bar{\bm{S}}_{.1}^{1/2} & \bm{0}_{t\times(N-t)}\\
(\bar{\bm{L}}_{2}^{\dagger})^{-1}\bar{\bm{S}}_{3}^{\dagger} & \bar{\bm{L}}_{2}
\end{bmatrix}\begin{bmatrix}\bm{G}_{1}^{\dagger} & \bm{0}_{t\times(N-t)}\\
\bm{G}_{2}^{\dagger} & \bm{G}_{3}^{\dagger}
\end{bmatrix}=\nonumber \\
\begin{bmatrix}\bm{S}_{.1}^{1/2} & \bm{0}_{t\times(N-t)}\\
(\bm{L}_{2}^{\dagger}\bm{U}_{1})^{-1}\bm{S}_{3}^{\dagger} & (\bm{U}_{1}^{\dagger}\bm{L}_{2})
\end{bmatrix}
\end{gather}
from which the following set of independent equations arises:
\begin{gather}
\begin{cases}
(i) & \bar{\bm{S}}_{.1}^{1/2}\,\bm{G}_{1}^{\dagger}=\bm{S}_{.1}^{1/2}\\
(ii) & (\bar{\bm{L}}_{2}^{\dagger})^{-1}\,\bar{\bm{S}}_{3}^{\dagger}\,\bm{G}_{1}^{\dagger}+\bar{\bm{L}}_{2}\,\bm{G}_{2}^{\dagger}=\,(\bm{L}_{2}^{\dagger}\,\bm{U}_{1})^{-1}\bm{S}_{3}^{\dagger}
\end{cases}\,
\end{gather}
which provides the ``completing'' solutions for matrix $\bm{G}$:
\begin{eqnarray}
\bm{G}_{1} & = & \bm{S}_{.1}^{1/2}\bar{\bm{S}}_{.1}^{-1/2}\label{eq: G_1 part (b)}\\
\bm{G}_{2}^{\dagger} & = & \bar{\bm{S}}_{2}^{-1}\left(\bm{G}_{3}^{-1}\bm{S}_{3}^{\dagger}-\bar{\bm{S}}_{3}^{\dagger}\bm{G}_{1}^{\dagger}\right)\label{eq: G_2 part (b)}
\end{eqnarray}
Up to now, we have shown how matrices $\bm{G}_{i}$ can be constructed.
Finally, it can be easily shown that matrix $\bm{F}_{1}$ should be
chosen as $\bm{F}_{1}=\bm{Z}_{1}-\sum_{i=1}^{3}\bm{G}_{1,i}\bar{\bm{Z}}_{i}$.This
concludes proof for $(\mathrm{b}$).

\section{MIS obtained by enforcing \protect \\
additional invariance in range-spread case \label{sec: Appendix _  Range-spread Raghavan}}

In this appendix we show that the statistic
\begin{gather}
\begin{cases}
\bm{x}_{1}\triangleq\mathrm{eig}(\bm{T}_{b})\\
\bm{x}_{2}\triangleq\mathrm{eig(}\bm{T}_{b}+\bm{T}_{a})
\end{cases}\label{eq: MIS_first_form}
\end{gather}
where $\bm{T}_{a}=(\bm{a}\bm{a}^{\dagger})$ ($\bm{a}\in\mathbb{C}^{M\times1}$)
and $\bm{T}_{b}\in\mathbb{H}^{M\times M}$, is a MIS for the elementary
action 
\begin{align}
\ell_{2,b}^{\star}(\bm{T}_{a},\bm{T}_{b})= & \left(\bm{U}_{d}^{\dagger}\,\bm{T}_{a}\,\bm{U}_{d},\,\bm{U}_{d}^{\dagger}\,\bm{T}_{b}\,\bm{U}_{d}\right)\,,\label{eq: elementary_op_range_spread}
\end{align}
where $\bm{U}_{d}\in\mathcal{U}(M)$. First, we observe that Eq. (\ref{eq: MIS_first_form})
is in one-to-one mapping with:
\begin{equation}
\begin{cases}
\bm{x}_{1}=\mathrm{eig}(\bm{T}_{b})\\
\bar{\bm{x}}_{2}\triangleq|\bm{k}|
\end{cases}\label{eq: MIS (form 2)}
\end{equation}
where $\bm{k}\triangleq(\bm{U}_{b}^{\dagger}\,\bm{a})$ and the modulus
$|\cdot|$ in Eq. (\ref{eq: MIS (form 2)}) should be intended element-wise.
Also, $\bm{U}_{b}$ denotes the eigenvector matrix of $\bm{T}_{b}$,
that is $\bm{T}_{b}=\bm{U}_{b}\,\bm{\Lambda}_{b}\,\bm{U}_{b}^{\dagger}$.
The existence of the aforementioned mapping can be proved as follows.
We start by observing that $\mathrm{eig(}\bm{T}_{b}+\bm{a}\bm{a}^{\dagger})$
can be obtained as the zeros (with respect to the variable $s$) of
the rational function \cite{Gu1994}:
\begin{equation}
w(s)=\left(1+\bm{k}^{\dagger}\,(\bm{\Lambda}_{b}-s\,\bm{I}_{M})^{-1}\,\bm{k}\right)\,.\label{eq: w(s) range_spread}
\end{equation}
 Also, since $(\bm{\Lambda}_{b}-s\bm{I}_{M})^{-1}$ is a diagonal
matrix, $w(s)$ depends only on $|\bm{k}|$. Therefore $\mathrm{eig(}\bm{T}_{b}+\bm{a}\bm{a}^{\dagger})$
can be obtained starting from $\bm{\Lambda}_{b}$ (viz. $\mathrm{eig}(\bm{T}_{b})$)
and $|\bm{k}|$. Vice versa, the vector $|\bm{x}|$ is obtained from
$\mathrm{eig(}\bm{T}_{b}+\bm{a}\bm{a}^{\dagger})$ and $\mathrm{eig(}\bm{T}_{b})$
by inverting Eq.~(\ref{eq: w(s) range_spread}), that is:
\begin{align}
\bm{k}^{\dagger}\,(\bm{\Lambda}_{b}-x_{2,i}\,\bm{I}_{M})^{-1}\,\bm{k} & =-1,\\
\sum_{n=1}^{M}\frac{|k_{n}|^{2}}{(\lambda_{b,n}-x_{2,i})} & =-1,\\
\bm{\alpha}_{i}^{T}\bm{\epsilon} & =-1,\quad i\in\{1,\ldots,M\}\label{eq: linear system}
\end{align}
where $\lambda_{b,n}$ is the $n$-th diagonal element of $\bm{\Lambda}_{b}$
and 
\begin{align}
\bm{\epsilon} & \triangleq\begin{bmatrix}|k_{1}|^{2} & \cdots & |k_{M}|^{2}\end{bmatrix}^{T},\\
\bm{\alpha}_{i} & \triangleq\begin{bmatrix}(\lambda_{b,1}-x_{2,i})^{-1} & \cdots & (\lambda_{b,N}-x_{2,i})^{-1}\end{bmatrix}^{T}.
\end{align}
It is shown hereinafter that the linear system in Eq. (\ref{eq: linear system})
(with respect to the unknown vector $\bm{\epsilon}$) admits a unique
solution. 

Indeed, the generic $\bm{\alpha}_{i}$ represents a scaled version
of $(\bm{E}_{p}\,\bm{v}_{a+b,i})$, where $\bm{E}_{p}\triangleq\mathrm{diag}\{\bm{\epsilon}\}$
and $\bm{v}_{a+b,i}$ denotes the $i$-th eigenvector of $\bm{T}_{b}+\bm{a}\bm{a}^{\dagger}$
\cite[Eq. (2.1)]{Gu1994}. However, since we assume that the eigenvalues
are distinct with probability one, the eigenvectors $\bm{v}_{b,i}$
will be linearly independent. Therefore, it follows that also the
set $\{\bm{\alpha}_{i}\}_{i=1}^{M}$ constitutes a linearly independent
basis. Such conclusion clearly implies that the system is invertible
and admits a unique solution; therefore there exists a one-to-one
correspondence between the statistics in Eq. (\ref{eq: MIS_first_form})
and (\ref{eq: MIS (form 2)}). 

Once established the correspondence between Eqs. (\ref{eq: MIS_first_form})
and (\ref{eq: MIS (form 2)}), it suffices to show that Eq. (\ref{eq: MIS (form 2)})
is a MIS for the group of trasformations specified in Eq. (\ref{eq: elementary_op_range_spread}).
In order to accomplish this task, we first prove \emph{invariance}
of statistic in Eq. (\ref{eq: MIS (form 2)}). Indeed, given the transformations:
\begin{gather}
\left\{ \widetilde{\bm{T}}_{a}=(\bm{U}_{d}^{\dagger}\,\bm{a}\bm{a}^{\dagger}\,\bm{U}_{d}),\quad\widetilde{\bm{T}}_{b}=(\bm{U}_{d}^{\dagger}\,\bm{T}_{b}\,\bm{U}_{d}),\right.
\end{gather}
It is readily shown that $\mathrm{eig}(\widetilde{\bm{T}}_{b})$ can
obtained as the zeros of:
\begin{gather}
\det(s\bm{I}-\bm{U}_{d}^{\dagger}\,\bm{T}_{b}\,\bm{U}_{d})=0\,\Leftrightarrow\det(s\bm{I}-\bm{T}_{b})=0
\end{gather}
thus coinciding with $\mathrm{eig}(\bm{T}_{b})$. Also, it holds
\begin{align}
|\tilde{\bm{U}}_{b}^{\dagger}\,\tilde{\bm{a}}| & =|\bm{U}_{b}^{\dagger}\,\bm{U}_{d}\,\bm{U}_{d}^{\dagger}\,\bm{a}|\nonumber \\
 & =|\bm{U}_{b}^{\dagger}\,\bm{a}|
\end{align}
Therefore the statistic in Eq. (\ref{eq: MIS (form 2)}) is invariant.
We then prove \emph{maximality}. Under the assumption
\begin{equation}
\begin{cases}
\mathrm{eig}(\bm{T}_{b})=\mathrm{eig}(\widetilde{\bm{T}}_{b})\\
|\bm{U}_{b}^{\dagger}\,\bm{a}|=|\widetilde{\bm{U}}_{b}^{\dagger}\,\widetilde{\bm{a}}|
\end{cases}\label{eq: max_assumption_range spread}
\end{equation}
it can be readily shown that there exists a unitary matrix $\bm{V}$
that ensures the equality $(\bm{V}^{\dagger}\bm{T}_{b}\bm{V})=\widetilde{\bm{T}}_{b}$,
namely $\bm{V}=(\bm{U}_{b}\,\bm{D}_{b}\,\widetilde{\bm{U}}_{b}^{\dagger})$,
where $\bm{D}_{b}$ is a diagonal matrix of arbitrary phasors. Similarly
we have employed the eigendecomposition $\widetilde{\bm{T}}_{b}=(\widetilde{\bm{U}}_{b}\,\widetilde{\bm{\Lambda}}_{b}\,\widetilde{\bm{U}}_{b}^{\dagger})$.
Additionally, in order to complete maximality proof, we need to prove
that the aforementioned transformation, when applied to $\bm{T}_{a}=\bm{a}\bm{a}^{\dagger}$,
can be adjusted to satisfy:
\begin{equation}
\bm{V}^{\dagger}(\bm{a}\bm{a}^{\dagger})\bm{V}=\tilde{\bm{a}}\tilde{\bm{a}}^{\dagger}
\end{equation}
After substitution, such condition can be rewritten as:
\begin{align}
\widetilde{\bm{U}}_{b}\,\bm{D}_{b}^{\dagger}\,\bm{U}_{b}^{\dagger}(\bm{a}\bm{a}^{\dagger})\bm{U}_{b}\,\bm{D}_{b}\,\widetilde{\bm{U}}_{b}^{\dagger} & =\tilde{\bm{a}}\tilde{\bm{a}}^{\dagger}\\
\bm{D}_{b}^{\dagger}\,\bm{U}_{b}^{\dagger}(\bm{a}\bm{a}^{\dagger})\bm{U}_{b}\,\bm{D}_{b} & =\widetilde{\bm{U}}_{b}^{\dagger}\tilde{\bm{a}}\tilde{\bm{a}}^{\dagger}\widetilde{\bm{U}}_{b}\\
\left[\bm{D}_{b}^{\dagger}\left(\bm{U}_{b}^{\dagger}\,\bm{a}\right)\right]\left[\bm{D}_{b}^{\dagger}\left(\bm{U}_{b}^{\dagger}\,\bm{a}\right)\right]^{\dagger} & =\left(\widetilde{\bm{U}}_{b}^{\dagger}\tilde{\bm{a}}\right)\left(\widetilde{\bm{U}}_{b}^{\dagger}\tilde{\bm{a}}\right)^{\dagger}
\end{align}
The above rank-one matrix equality can be achieved by enforcing the
vector equality
\begin{equation}
\left[\bm{D}_{b}^{\dagger}\left(\bm{U}_{b}^{\dagger}\,\bm{a}\right)\right]=\left(\widetilde{\bm{U}}_{b}^{\dagger}\tilde{\bm{a}}\right)
\end{equation}
by choosing each element of the diagonal matrix $\bm{D}_{b}^{\dagger}$
in order to rotate each phase term of ($\bm{U}_{b}^{\dagger}\,\bm{a}$)
aiming at imposing $\angle\left(\bm{U}_{b}^{\dagger}\,\bm{a}\right)=\angle\left(\widetilde{\bm{U}}_{b}^{\dagger}\tilde{\bm{a}}\right)$,
since $|\bm{U}_{b}^{\dagger}\,\bm{a}|=|\widetilde{\bm{U}}_{b}^{\dagger}\tilde{\bm{a}}|$
by definition (cf. Eq. (\ref{eq: max_assumption_range spread})).
Therefore Eq. (\ref{eq: MIS (form 2)}) (resp. Eq. (\ref{eq: MIS_first_form}))
is a MIS for the aforementioned group of trasformations. 

\bibliographystyle{IEEEtran}
\bibliography{IEEEabrv,bib_adapt_det}

\end{document}